%% file: paper.tex
\documentclass[compsoc,conference,a4paper,10pt,times]{IEEEtran}
\IEEEoverridecommandlockouts
\usepackage{cite}
\usepackage{amsmath,amssymb,amsfonts}
\usepackage{algorithmic}
\usepackage{graphicx}
\usepackage{textcomp}
\usepackage{bmpsize}
\usepackage{xcolor}
\usepackage{lipsum}
\def\BibTeX{{\rm B\kern-.05em{\sc i\kern-.025em b}\kern-.08em
    T\kern-.1667em\lower.7ex\hbox{E}\kern-.125emX}}

\usepackage{xprintlen}
\usepackage{subcaption}
\usepackage{colortbl}
\usepackage{ marvosym }
\usepackage{fdsymbol}
\usepackage{manfnt}
\usepackage{fontawesome5}
\usepackage{ragged2e}

\input{include}

\newcommand{\nobroadcast}{\faLocationArrow}
\newcommand{\scalability}{\faCompress}
\newcommand{\verificationdelay}{\faSpinner}
\newcommand{\desync}{\faHistory}

\newcommand{\replay}{\faRedo*}
\newcommand{\disconnect}{\faPowerOff}
\newcommand{\masquerade}{\faMask}

\newcommand{\detectiondelay}{\faBellSlash}

\newcommand{\storage}{\faDatabase}
\newcommand{\communication}{\faMailBulk}
\newcommand{\delay}{\faHourglassHalf}

\newcommand{\full}{$\medblackdiamond$}
\newcommand{\part}{$\meddiamond$}

\usepackage[colorlinks=true,urlcolor=black]{hyperref}

\usepackage{etoolbox}
\makeatletter
\patchcmd{\hyper@makecurrent}{%
    \ifx\Hy@param\Hy@chapterstring
        \let\Hy@param\Hy@chapapp
    \fi
}{%
    \iftoggle{inappendix}{%true-branch
        \@checkappendixparam{chapter}%
        \@checkappendixparam{section}%
        \@checkappendixparam{subsection}%
        \@checkappendixparam{subsubsection}%
        \@checkappendixparam{paragraph}%
        \@checkappendixparam{subparagraph}%
    }{}%
}{}{\errmessage{failed to patch}}

\newcommand*{\@checkappendixparam}[1]{%
    \def\@checkappendixparamtmp{#1}%
    \ifx\Hy@param\@checkappendixparamtmp
        \let\Hy@param\Hy@appendixstring
    \fi
}
\makeatletter

\newtoggle{inappendix}
\togglefalse{inappendix}

\apptocmd{\appendices}{\toggletrue{inappendix}}{}{\errmessage{failed to patch}}
\usepackage[nolist]{acronym}
\begin{acronym}[AAAAAAA]	
	\acro{ISP}{Internet Service Provider}
    \acro{ACK}{acknowledgement}
    \acro{CAN}{Controller Area Network}
    \acro{CRC}{cyclic redundancy check}
    \acro{CSMA}{carrier-sense multiple access}
    \acro{DLC}{data length code}
    \acro{DLL}{Data Link Layer}
    \acro{DoS}{Denial of Service}
    \acro{ECU}{Electronic Control Unit}
    \acro{EOF}{end-of-frame}
    \acro{ID}{identifier}
    \acro{IDS}{Intrusion Detection System}
    \acro{MAC}{Message Authentication Code}
    \acro{NRZ}{non-return-to-zero}
    \acro{SOF}{start-of-frame}
    \acro{RBF}{Reactive Bit Flipping}
    \acro{IDS}{Intrusion Detection System}
    \acro{SIP}{Semiconductor Intellectual Property}

	\acrodefplural{ISP}[ISPs]{Internet Service Provider}
\end{acronym}

\newcommand{\name}{\textsc{Caiba}\xspace} % On-the-fly MAC Overwrite
\newcommand{\fullname}{Compact And Instantaneous Bus Authentication\xspace}%{Compact And Instantaneous \fix{Broadcast/Bus} Authentication\xspace} % On-the-fly MAC Overwrite

\newcommand{\kgroup}{$k^{\text{group}}$\xspace} % On-the-fly MAC Overwrite
\newcommand{\ksource}{$k^{\text{source}}_{j}$\xspace} % On-the-fly MAC Overwrite

\begin{document}
%-------------------------------------------------------------------------------

\title{\textsc{Caiba}: Multicast Source Authentication for CAN Through Reactive Bit Flipping}

\author{
{\rm Eric Wagner$^{\ast\dagger}$, Frederik Basels$^{\ast}$, Jan Bauer$^{\ast}$, Till Zimmermann$^{\ddagger}$,  Klaus Wehrle$^{\dagger}$, and Martin Henze$^{\mathsection\ast}$} \\
    {\rm $^\ast$\textit{Cyber Analysis \& Defense}, Fraunhofer FKIE $\cdot$ \{firstname.lastname\}@fkie.fraunhofer.de } \\
    {\rm $^\dagger$\textit{Communication and Distributed Systems}, RWTH Aachen University $\cdot$ \{lastname\}@comsys.rwth-aachen.de} \\
    {\rm $^\ddagger$\textit{Distributed Systems Group}, Osnabrück University $\cdot$ zimmermann@uos.de} \\
    {\rm $^\mathsection$\textit{Security and Privacy in Industrial Cooperation}, RWTH Aachen University $\cdot$ henze@spice.rwth-aachen.de}
}

\maketitle
\thispagestyle{plain}
\pagestyle{plain}

\begin{abstract}

    \acp{CAN} are the backbone for reliable intra-vehicular communication.
    Recent cyberattacks have, however, exposed the weaknesses of \ac{CAN}, which was designed without any security considerations in the 1980s.
    Current efforts to retrofit security via intrusion detection or message authentication codes are insufficient to fully secure \ac{CAN} as they cannot adequately protect against masquerading attacks, where a compromised communication device, a so-called electronic control units, imitates another device.
    To remedy this situation, multicast source authentication is required to reliably identify the senders of messages.
    In this paper, we present \name, a novel multicast source authentication scheme specifically designed for communication buses like \ac{CAN}.
    \name relies on an authenticator overwriting authentication tags on-the-fly, such that a receiver only reads a valid tag if not only the integrity of a message but also its source can be verified.
    To integrate \name into \ac{CAN}, we devise a special message authentication scheme and a reactive bit overwriting mechanism.
    We achieve interoperability with legacy \ac{CAN} devices, while protecting receivers implementing the AUTOSAR SecOC standard against masquerading attacks without communication overhead or verification delays.
\end{abstract}

\begin{IEEEkeywords}
    multicast source authentication, CAN bus, message authentication codes, AUTOSAR SecOC
\end{IEEEkeywords}

\acresetall

\input{content.tex}

\balance
\bibliographystyle{IEEEtranS}
\bibliography{references}

\section*{Data Availability}

Our modifications to the SDCC~\cite{2019_cena_sdcc} to support \name are available at \url{https://github.com/fkie-cad/caiba}.

\appendix

\renewcommand{\thesubsection}{\Alph{subsection}}

\renewcommand\thesubsectiondis{\Alph{subsection}.}

\renewcommand\thesubsubsection{\thesubsection.\arabic{subsubsection}}

\renewcommand\thesubsubsectiondis{\thesubsection.\arabic{subsubsection}.}

\subsection{Detailed Analysis of Selected Schemes}
\label{app:cryptanalysis}

A recent survey on CAN authentication schemes~\cite{2024_lotto_survey} revealed weaknesses with many of the proposed CAN authentication schemes but classified others as \textit{secure protocols}~(LiBrA-CAN~\cite{2012_groza_libra}, LinAuth~\cite{2012_lin_cyber}, LCAP~\cite{2012_hazem_lcap}, CaCAN~\cite{2014_kurachi_cacan}, and AuthentiCAN~\cite{2020_marasco_authentican}).
With regard to the eleven drawbacks defined in \autoref{ssec:rw}, we took a closer look at the latter schemes~(\cf \autoref{app:cryptanalysis:1}\,--\,\ref{app:cryptanalysis:5}), as well as some additional authentication schemes not considered by the authors~(\ie Watermarking~\cite{2022_michaels_watermarking}, CANTO~\cite{2020_groza_canto}, ZBCAN~\cite{2023_serag_zbcan}, CAN-MM~\cite{2024_oberti_canmm}, LEAP~\cite{2019_lu_leap}, and CAN-TORO~\cite{2020_groza_highly} in \autoref{app:cryptanalysis:6}\,--\,\ref{app:cryptanalysis:11}), and re-assess their security in this section.
Overall, we find that none of the current schemes offer full protection~(as shown in \autoref{tab:related-work}), hence justifying the existence of \name.

\subsubsection{LiBrA-CAN}
\label{app:cryptanalysis:1}
\hfill\vspace{1mm}

\textbf{Description.}
LiBrA-CAN~\cite{2012_groza_libra} uses a Mixed MAC approach for integrity protection, \ie keys are shared among a subset of nodes and each node can only verify a fraction of each authentication tag.
These keys are initially distributed by computationally superior master node.
To authenticate a frame sent to the bus, the sending ECU first authenticates it to the master node or an helper node.
This additional node knows additional keys to compute additional authentication data such that the remaining ECUs can eventually~(more than one helper node may be required) verify the integrity of the original message.
\vspace{1mm}

\textbf{Analysis.} 
LiBrA-CAN introduces verification delays and communication overhead, similarly to TESLA-based protocols~\cite{2000_perrig_tesla,2001_perrig_tesla2}~(\cf~\autoref{sec:caiba:stateoftheart}).
This overhead grows linearly with the number of nodes and, for a certain number of nodes, the use of digital signatures may even become beneficial over using LiBrA-CAN~\cite{2012_groza_libra}.
Additionally, a master node with knowledge of all secret keys exposes a single point of failure, as an attacker that compromises this node gains full control over the network.

\vspace{1mm}
\textbf{Takeaway.}
LiBrA-CAN's scalability is limited~\scalability, the authentication requires the sending of additional frames~\communication~that lead to a verification delay~\verificationdelay , and the master node offers an attractive target to attackers as it is in possession of all keys~\masquerade.

\subsubsection{LinAuth}
\hfill\vspace{1mm}

\textbf{Description.}
In LinAuth~\cite{2012_lin_cyber}, each ECU pair shares a secret key.
For each CAN~ID an ECU transmits or receives, the ECU additionally needs to know the group of other ECUs that are also interested messages with this ID.
A sender of a message computes an authentication tag for each interested receiving ECU.
The available space is then evenly split among each receiver and fills with the truncated authentication tags as well as the least-significant bits of each ECU-specific counter.
Each receiving ECU knows where its authentication data starts and ends within a frame and can thus verify the truncated authentication tag.
\vspace{1mm}

\textbf{Analysis.} 
LinAuth requires significant configuration and storage to track which node is interest in which CAN~IDs.
Moreover, LinAuth does not scale to realistic network sizes.
Reserving \SI{24}{\bit} for authentication data and transmitted \SI{4}{\bit} for each counter, as proposed by the \code{SecOC Profile\,3\,(JASPAR)} of AUTOSAR~\cite{autosar}, already leaves no space for authentication data with 6 receiving ECU.
\vspace{1mm}

\textbf{Takeaway.}
LinAuth requires significant upfront configuration and additional storage~\storage~and does not scale to realistic network sizes regarding the number of ECUs that are interested in a given CAN~ID~\scalability~as space in each frame must be reserved for each receiver~\communication.

\subsubsection{LCAP}
\hfill\vspace{1mm}

\textbf{Description.}
In LCAP~\cite{2012_hazem_lcap}, CAN frames are authenticated by including a 2-byte ``magic number'' in the payload.
This magic number is an element of a hash chain only known to the sending ECUs of a given communication group, while the receiving ECUs know the previous element of this chain.
Thus, upon reception of a frame, receivers can verify that the magic number indeed stems from the alleged sender.
\vspace{1mm}

\textbf{Analysis.} 
In LCAP, all communication groups must be known in advance and the sender in each such group must precompute and store a hash chain of a significant length, amounting to several kB of data per group.
If all elements of a chain are consumed, a new chain must be computed and handshake messages must be exchanged on the bus, leading to protocol and communication overhead.
Most importantly, LCAP is vulnerable to the hijacking of magic numbers.
By decoding the magic number of a payload while immediately jamming the bus afterwards, an attacker could  ensuring the frame is rejected by other ECUs.
Before the original frame is retransmitted by the sender, the attacker can now exploit this~(still valid) magic number to inject allegedly legitimate messages.
 
%Similarly, an attack could (reactively) overwrite a payload while it is transmitted, skipping over the magic number, as proposed in \cite{2023_yousik_bit_modification} and refined in this paper~(\cf~\autoref{sec:bitflip}).
%\vspace{1mm}

\textbf{Takeaway.} LCAP produces storage~\storage~and comunication~\communication~overhead and is vulnerable to frame interceptions~\replay.

\subsubsection{CaCAN}
\hfill\vspace{1mm}

\textbf{Description.}
Similarly to \name, CaCAN~\cite{2014_kurachi_cacan} relies on an authenticator node to assist in authenticating CAN frames.
This authenticator shares a secret key with each ECUs and has the capability of overwrite in-transmission frames with an error frame, ensuring that the frame is not received by other ECUs.
A sending ECU computes a MAC for the message including a counter and transmits the message, the least-significant bits of the counter, and the first byte of the MAC in a payload of a CAN frame.
The authenticator verifies this MAC and disrupts the message if the verification fails.
\vspace{1mm}

\textbf{Analysis.} 
In CaCAN no mechanism for resynchronization for the counter after lost frames is devised.
Concerning vulnerabilities, CaCAN's authenticator has access to all keys, an attack does not need to compromise it to circumvent CaCAN's protection.
It is sufficient to disconnect the authenticator from bus, to disable all authenticity verification without anyone noticing.
Afterwards, the attacker can masquerade as any ECU through an attached or compromised ECU as with legacy CAN networks.
\vspace{1mm}

\textbf{Takeaway.}
Disconnecting CaCAN's authenticator disable all security measures~\disconnect. 
Additionally, the authenticator may be compromised to gain access to all key and masquerade as other devices~\masquerade.
Finally, frame loss leads to desynchronizations of counters that cannot be recovered from~\desync.

\subsubsection{MAuth-CAN}
\hfill\vspace{1mm}

\textbf{Description.}
MAuth-CAN~\cite{2020_jo_mauth} also relies on an authenticator node to authenticate CAN frames.
For each CAN ID, the sending ECU and authenticator establish a session key.
With this key, a 32-bit authentication tag is generated and included in each CAN frame, split into the extended identifier and payload fields.
The authenticator verifies this tag and broadcasts a report only if the verification fails.
Receiving ECUs thus wait for a predefined time after frame detection for this report and only process the message when no report got received.
If a report is received, its authenticity if verified by the ECU and the frame it reports on is discarded.
\vspace{1mm}

\textbf{Analysis.}
The MAuth-CAN authenticator knows all keys and if it is disconnected, no frame verification takes place and no reports are sent.
The receiving ECUs can, however, not differentiate if reports are not sent because the frame is authentic or because the authenticator is not operational.
Thus, after disconnecting the authenticator, the attack can masquerade as any ECU through an attached or compromised ECU.
Even without attacks, MAuth-CAN introduces a delay for all frames as ECUs wait for a potential report, and resynchronization requires to perform new session key exchange.
\vspace{1mm}

\textbf{Takeaway.}
Disconnecting the authenticator disable all security measures~\disconnect. 
Then, or if the authenticator is compromised to gain access to all key and masquerade as other devices~\masquerade.
Moreover, all frames are buffered by the receiver~\verificationdelay~and resynchronizations is expensive~\desync.

\subsubsection{AuthentiCAN}
\label{app:cryptanalysis:5}
\hfill\vspace{1mm}

\textbf{Description.}
In AuthentiCAN~\cite{2020_marasco_authentican}, a nonce list is exchanged between each ECU pair, secured by asymmetric encryption based on a broadcasted public key.
The payload of each frame then consists of the encrypted concatenation of a message and the first yet unused nonce from the list.
The receiver of a message can decrypt a message and verify that the transmitted nonce matches the expected nonce, which allegedly verifies the authenticity of a message.
\vspace{1mm}

\textbf{Analysis.} 
AuthentiCAN introduces significant overhead \wrt to memory and bandwidth consumption to exchange and keep track of the nonce lists and consequently primarily addresses CAN-FD with its higher data rate.
Secondly, AuthentiCAN only protects the communication between two ECUs and does not support broadcast communication, \ie frames intended for multiple receivers.

\vspace{1mm}
\textbf{Takeaway.}
AuthentiCAN does not support broadcast communication~\nobroadcast~and causes significant memory~\storage~and communication~\communication~overhead.

\subsubsection{Watermarking}
\label{app:cryptanalysis:6}
\hfill\vspace{1mm}

\textbf{Description.} 
The idea of watermarking~\cite{2022_michaels_watermarking} is to overlay a high-frequency signal over ordinary CAN transmissions.
These overlayed signals transmit a time-varying watermark, generated by a random number generator that is seeded with a key known to all legitimate devices connected to the bus.
\vspace{1mm}

\textbf{Takeaway.} A compromised ECU can transmit messages with legitimate watermarks, enabling the attack to masquerade as any other device~\masquerade.

\subsubsection{CANTO}
\label{app:cryptanalysis:6}
\hfill\vspace{1mm}

\textbf{Description.} 
CANTO~\cite{2020_groza_canto} provides frame authentication through covert timing channels.
It assumes that all CAN~IDs are transmitted in a regular pattern that is known in advance by all receivers, which is assisted by an optimal a priori frame scheduling to maximize inter-frame spacing and avoid collisions.
At the scheduled time, a sending ECU computes a MAC of the frame with a shared group key and uses it to derives an additional delay for the frame.
The receiving ECUs then verify that this delay from the expected arrival time matches the expectation for the transmitted frame.
\vspace{1mm}

\textbf{Analysis.} Because of its assumption that traffic is sent in repeating patterns, the real-world applicability of CANTO to protect in-vehicule communication may be limited due to spontaneous actions, \eg by humans in the loop.
Moreover, even if CAN traffic exhibits the required regularity, the deployment of CANTO requires the modification of each ECU to support the modified scheduling.
Finally, CANTO relies on a shared group key, thus an attack that compromises one ECU has access to this key and can masquerade as any other ECU.
\vspace{1mm}

\textbf{Takeaway.} CANTO does not protect against a single compromised ECU~\masquerade~and reschedules the transmission of frames~\delay, ultimately limiting  deployability in vehicles as all ECUs must support CANTO.

\subsubsection{ZBCAN}
\hfill\vspace{1mm}

\textbf{Description.} 
In ZBCAN~\cite{2023_serag_zbcan}, each ECU shares a pairwise secret with an authenticator node.
Transmissions do not occur instantaneously, but instead are executed at specific time slots.
Therefore, the time after the last transmission is split into discrete time spans, each consisting of a predefined number of time slots.
The sending ECU then computes its time slot based on the CAN~ID, the secret shared with the authenticator, and an implicit message counter.
For a given CAN~ID, only a subset of all time slots are available to divide them into priority classes.
The authenticator monitors the network and verifies the time slots of each message based on its ID, and overwrite mis-timed messages with an error frame, such that only verified messages are received by the other ECUs.
\vspace{1mm}

\textbf{Analysis.}
While ZBCAN does not consume any additional bandwidth, it is also susceptible to several %problems and
weaknesses.
First, the priority of IDs can be inverted: Consider a high priority message that is generated just after its time slot has passed.
Then, a lower priority message would take precendence within this time span before the high priority message has another chance to be transmitted, delaying potentially critical messages.
Secondly, with the proposed time span length of 64~nominal bit times for each priority class ZBCAN only achieves the equivalent security of a 6-bit authentication tag.
Thirdly, the authenticator can be simply disconnected from the bus as all messages~(spoofed or not) are accepted by the network without an authenticator.
Finally, ZBCAN only authenticates the intention of the sender to transmit, but not the content of the frame.
Hence, an attacker may wait for a transmission to overwrite its content~(as shown in~\autoref{sec:bitflip}) and the resulting error frame to compromise the channel.
\vspace{1mm}

\textbf{Takeaway.}
ZBCAN causes some transmission delays~\delay, only protects the sender intention to send but not the content of this transmission with relatively low security levels~\replay, and does not protect against disconnecting the authenticator from the bus~\disconnect.

\subsubsection{CAN-MM}
\hfill\vspace{1mm}

\textbf{Description.}
With CAN-MM~\cite{2024_oberti_canmm}, ECUs can be incrementally retrofitted with transmitter and receiver modules.
These modules compute a MAC based on a frame's content and a group key and then multiplex the transmission of the CAN frame and the MAC with On-Off keying.
A receiving ECU equipped with the receiver module can de-mulitplex the transmission and verify the MAC, while all other ECUs decode ordinary CAN frames.
\vspace{1mm}

\textbf{Takeaway.}
CAN-MM does not protect against masquerading attacks by compromised ECU because of its reliance on group keys~\masquerade.

\subsubsection{LEAP}
\hfill\vspace{1mm}

\textbf{Description.}
In LEAP~\cite{2019_lu_leap}, each pair of ECUs share a secret key which is used to compute a keystream by encrypting the CAN~ID of a message.
The first eleven bits of this keystream are embedded into the payload at a location determined by next bits of the keystream such that an eavesdropper does not know which payload bits consist of the authentication tag.
Finally, the remaining bits of the keystream are used to encrypt the payload.
The intended receiver of a message can perform these steps in reverse order to decrypt the payload and verify the correct embedding.
\vspace{1mm}

\textbf{Takeaway.}
LEAP does not support broadcast communication~\nobroadcast, and requires relatively high amounts of storage for key material~\storage~to compute authentication tags embedded into the payload~\communication.

\subsubsection{CAN-TORO}
\label{app:cryptanalysis:11}
\hfill\vspace{1mm}

\textbf{Description.} 
CAN-TORO~\cite{2020_groza_highly} proposes to encrypt CAN~IDs with order-preserving encryption to hide the sender from eavesdroppers and to authenticate them without interfering with message prioritization.
Each legitimate ECU keeps track of a mapping of IDs to encrypted IDs which is derived from a group key and updated regularly, \eg once a second.
Received CAN frames are discarded if they contain an invalid ID, where the valid IDs change constantly.
\vspace{1mm}

\textbf{Analysis.} 
For CAN-TORO, all ECUs software needs to be modified, as individual ECUs not supporting the protocol interferes with prioritization and may lead to the discarding of valid CAN frames.
Moreover, an attack has a short period of time to replay a valid ID before the mapping changes or to outright overwrite the payload of a frame.
Finally, by compromising a single ECU, an attacker gains access to the group key and can then masquerade as any other ECU. 
\vspace{1mm}

\textbf{Takeaway.} 
CAN-TORO is susceptible to ID reuse by frame interception~\replay~and does not protect against masquerading attacks by compromised ECUs~\masquerade. 
Moreover, tracking the mapping of encrypted IDs cost valuable storage~\storage.

\subsubsection{Other Approaches}
\hfill\vspace{1mm}

All approaches not further analyzed here~(CANAuth~\cite{2011_van_canauth}, Car2X~\cite{2011_schweppe_car2x}, Woo-Auth~\cite{2014_woo_practical}, VeCure~\cite{2014_wang_vecure}, LeiA~\cite{2016_radu_leia}, vatiCAN~\cite{2016_nurnberger_vatican}, VulCAN~\cite{2017_vanbulck_vulcan}, TOUCAN~\cite{2019_bella_toucan}, and S2-CAN~\cite{2021_pese_s2can}) have their weaknesses discussed in detail by Lotto~\etal~\cite{2024_lotto_survey}.

\subsection{Security Proof}
\label{app:proof}

In this section, we want to formalize the security of \name.
\name's security relies on two key assumptions:
\begin{itemize}
    \item An \emph{adversary} $\mathcal{A}$ cannot simultaneously compromise an ECU and the authenticator.
    \item $\mathcal{A}$  cannot undetectably overwrite bits, unless they compromise the authenticator.
\end{itemize}
From these two assumptions, we can proof that \name achieves the same security levels as an AUTOSAR SecOC instance with the same tag length.
Here, we assume that the underlying MAC schemes are deterministic ($m$ has exactly one valid $t$) and ideal (an adversary best strategy is to guess a valid tag with a success rate of $1/2^{|t|}$, where $|t|$ is the bit-length of $t$).
Thus, \eg 3-byte long tags lead to a 1 in $2^{24}$ chance of a tag being misclassified as valid, \ie a security level of 24 bit, for \name and AUTOSAR SecOC alike. 

We proof the security of \name with a game as typically done for MAC schemes~\cite{2020_boneh_graduate}:
$\mathcal{A}$ may query an oracle with messages $m_i\in \mathcal{M}$ for $t_i$ and eventually outputs a candidate forgery $(m^\prime,t^\prime), m^\prime\notin \mathcal{M}$, where $\mathcal{M}$ is the message space. 
$\mathcal{A}$ wins this game if the tag $t^\prime$ is valid for the message $m^\prime$. 
The security of the scheme is then expressed as P$[\mathcal{A}$ wins$]$, i.e., the probability that $\mathcal{A}$ wins this game. 

We have to consider two cases. 
First, an adversary may have compromised an ECU.
In this case, $\mathcal{A}$ can generate $t^i$ but not $t^s$ (unless the compromised ECU is authorized to send CAN ID).
$\mathcal{A}$ may query the oracle for $t^s_i$ or $t_i$ for $m_i \in \mathcal{M}$.
However, to interfere $t^{s\prime}$, $\mathcal{A}$ has no better strategy than guessing as the underlying MAC scheme is considered secure.
Meanwhile, there exists no better strategy than guessing $t'$ either, as otherwise the MAC scheme to compute $t^s$ were not ideal.
As the receiver expects to receive $t^{i\prime}$, \ie $t$ must be $t^i \oplus t^s$ before modification, $\mathcal{A}$'s best strategy is to randomly guess a tag, i.e., P$[\mathcal{A}$ wins$] =1/2^{|t|}$.

Secondly, an adversary may have compromised the authenticator.
In this case, $\mathcal{A}$ knows the keys to compute all source-authenticating tags $t^s$, but cannot compute integrate-protecting tags $t^i$.
$\mathcal{A}$ could inject a frame without overwriting it.
However, therefore they would need to guess a valid $t^i$, which only succeeds with a probability of $1/2^{|t^i|}$.
Alternatively, $\mathcal{A}$ could modify a transmitted message to modify its origin (CAN ID) or content.
However, if any of these two fields are modified, $t^i$ is no longer valid and $\mathcal{A}$ would need to guess a new valid tag.

In both cases, P[$\mathcal{A}$ wins$]=1/2^{|t|}$. Thus, the security of \name depends on $|t|$. In practice, MAC schemes are not ideal, so the MAC scheme chosen in a concrete deployment will cause marginally lower security.

\end{document}

%% file: include.tex
\usepackage{xcolor}
\usepackage[detect-weight=true,detect-family=true]{siunitx}
\usepackage{xspace}

\usepackage{tikz}
\usetikzlibrary{positioning}

\usepackage{multirow}

\usepackage{textgreek}

\definecolor{darkgrey}{RGB}{80,80,80}
\definecolor{lightgrey}{RGB}{170,170,170}

\definecolor{brown}{HTML}{a52a2a}
\definecolor{darkcyan}{HTML}{0a888a}

\definecolor{sender}{RGB}{132,200,234}
\definecolor{receiver}{RGB}{62,206,90}
\definecolor{authenticator}{RGB}{241,177,69}

\usepackage{adjustbox}
\usepackage{booktabs}
\usepackage{wasysym}
\usepackage{tabularx} % colored table background
\usepackage{multirow}
\usepackage{arydshln} % includes hdashline and cdashline
\usepackage{siunitx}
\usepackage{enumitem}

\usepackage{balance}

\newcommand\Tstrut{\rule{0pt}{8pt}}         % = `top' strut
\newcommand\Bstrut{\rule[-3pt]{0pt}{6pt}}

\newcommand{\etal}{\textit{et~al.}\xspace}
\newcommand{\ie}{\textit{i.e.},\xspace}
\newcommand{\eg}{\textit{e.g.},\xspace}
\newcommand{\cf}{\textit{cf.}\xspace}

\newcommand{\wrt}{w.r.t.\xspace}

\newcommand{\code}[1]{\texttt{#1}}

\newcommand{\CUT}[1]{{\iffalse#1\fi}}

%% file: content.tex
\begin{figure*}
    \centering
    \includegraphics[width=\textwidth]{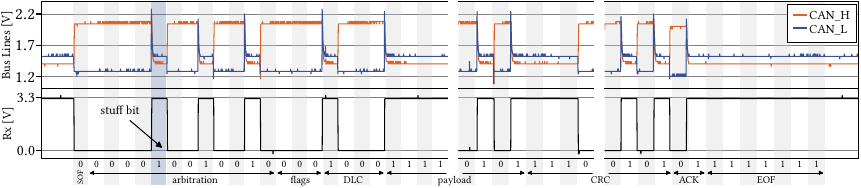}

    \caption{
    %    \vspace*{-0.2cm}
        To transmit data, CAN uses differential encoding on two wires, CAN\_H and CAN\_L. We display a base format CAN frame with 8\,bytes of payload as used to control hunreds of millions of cars every day. Note the gap in the payload and CRC field for increased readability.
    }
    \label{fig:can_frame}
\end{figure*}

\section{Introduction}

Virtually all motor vehicles currently on the roads are equipped with hundreds of small embedded computers, so-called \acp{ECU}, that monitor and control vital vehicle functions~\cite{kim2020autosar}.
To realize overarching functionality, these \acp{ECU} require means to communicate reliably with each other in harsh environments.
The de-facto standard for such in-vehicle interconnection is the \ac{CAN} bus.
Developed in the 1980s under the assumption of vehicles being physically isolated systems, security was not a design goal of \ac{CAN}. %. where the air-gap offers sufficient protection against attacks

However, contrary to this original assumption, modern cars offer more and more (wireless) connectivity and are thus increasingly exposed to cyberthreats %from cyberattacks 
with potentially 
fatal consequences~\cite{2011_checkoway_comprehensive}.
For example, a remote compromise of the infotainment system of a Tesla Model~S enabled attackers to  control the car's acceleration, leaving potential passengers at the attacker's mercy~\cite{2017_nie_freefall}.
This example is not an isolated incident, but only one of many recent attack demonstrations~\cite{2011_checkoway_comprehensive, 2017_nie_freefall, 2015_miller_remote, 2018_nie_over, 2020_wen_plug, 2015_foster_fast,2022_de_canflict , 2024_car_theft}.
%These attacks exploit that, once an individual \ac{ECU} has been compromised or spoofed, \ac{CAN} provides no protection %such that messages can be injected or manipulated at will~\cite{2010_koscher_experimental}.
%against this \ac{ECU} masquerading as other, potentially much more critical, entities and injecting arbitrary messages in their name~\cite{2010_koscher_experimental}.
Those attacks exploit that, once an \ac{ECU} has been compromised, \ac{CAN} provides no protection against \ac{ECU} masquerading and arbitrary message spoofing~\cite{2010_koscher_experimental}.
Concerningly, the dark web offers car theft devices today that exploit the vulnerability of CAN~\cite{2024_car_theft}.
These devices are actively used by criminals to steal modern cars~\cite{2024_car_theft}.

% Despite its vulnerabilities, the replacement of CAN is hardly imaginable.
% CAN is used by hundreds of \acp{ECU} in each car, stemming from different manufactures, which makes it hard to coordinate the use of an entirely new protocol.
% Moreover, CAN has a track record of providing reliable communication in harsh environments and a suitable fully secure replacement simply does not exist.

To overcome these serious vulnerabilities, different streams of research propose \emph{intrusion detection}, \emph{covert channels}, as well as \emph{cryptographic approaches}~\cite{2021_aliwa_cyberattacks}.
While \acp{IDS}~\cite{2022_limbasiya_systematic} do not interfere with legacy devices, they provide imperfect detection and often only alert after a sequence of manipulations~\cite{2023_serag_zbcan}.
Moreover, recent research has shown how these \acp{IDS} can be evaded by an attacker in practice~\cite{2021_bhatia_evading,2018_sagong_cloaking}.
Covert channels~\cite{2022_michaels_watermarking, 2020_groza_canto , 2024_oberti_canmm,2023_serag_zbcan} only provide reduced security and cannot protect against masquerading attacks.
Cryptographic approaches~\cite{2024_lotto_survey}, also comprising the  AUTOSAR  specification for Secure Onboard Communication (SecOC)~\cite{autosar} currently in adoption by multiple vendors~\cite{kim2020autosar}, mostly rely on group-keys to protect against outside attackers. 
AUTOSAR SecOC reserves a fraction of each CAN payload, \eg \SI{28}{\bit}, to transmit an authentication tag computed with the help of a secret key shared among all \acp{ECU}.
These approaches thus do not protect against compromised \acp{ECU}~(as used during \eg the Tesla Model~S attack~\cite{2017_nie_freefall}) as those can still masquerade as any entities they can receive from.

%\mh{the AUTOSAR specification also mentions asymmetric cryptography, where this argument might not hold}
% From Autosar: "In case of an asymmetric approach using digital signatures instead of the MAC-approach described throughout the whole document, some adaptations must be made". So they don't seriously consider digital signatures

%In 2016, Nie~\etal remotely compromised the infotainment system of a Tesla Model~S~\cite{2017_nie_freefall}.
%Gaining root access then allowed the attackers to \eg control the car's break and acceleration, leaving potential passengers at their mercy.
%Especially modern cars that offer more and more (wireless) connectivity are increasingly exposed to threats from cyberattacks with potentially fatal consequences.
%All these attacks are assisted by an insecure CAN bus, the standard for reliable in-car communication between \acp{ECU}.

%The CAN bus was designed for reliability but security was not in its design consideration.
%Cars and other systems controlled by CAN were considered physically isolated system where the air-gap offers sufficient protection against attacks.
%However, the recent attacks have shown that individual \acp{ECU} can be compromised or spoofed, such that access to CAN, which offers no protection against messages injection or manipulation, is gained.

The main challenge underlying all these approaches is the need for both reliable and immediate protection of multicast communication.
Most notably, CAN is a broadcast communication protocol, where messages are often intended for multiple receivers~\cite{2012_groza_libra}.
Typically, to verify the source of a message in a multicast scenario, digital signatures are used as authentication and verification rely on different keys.
However, digital signatures are not applicable in CAN due to excessive computational and bandwidth requirements~\cite{2021_aliwa_cyberattacks}.
CAN frames hold at most \SI{8}{bytes} of payload and even the up to \SI{64}{bytes} of the CAN-FD extension are insufficient to support asymmetric cryptography.
An alternative for source authentication in multicast communication relying on more resource-conscious symmetric cryptography is TESLA~\cite{2000_perrig_tesla}.
TESLA takes advantage of delayed key releases which, however, leads to verification delays and is thus not applicable to the safety-critical operation of CAN~\cite{2021_aliwa_cyberattacks}.
To fully protect CAN against the potentially fatal consequences of a compromised \ac{ECU}, a multicast source authentication scheme is needed that only occupies a few bytes per message, yet delivers instantaneous source verification.

In this paper, we first address the general lack of adequate multicast source authentication schemes by proposing \fullname~(\name).
To the best of our knowledge, \name is the first multicast source authentication scheme without verification delay while relying on tag sizes as small as traditional \acp{MAC} %schemes 
with the same security strength.
\name can cryptographically protect any network against masquerading attacks as long as it is possible to place a node that can listen to and overwrite all communications.
In a nutshell, \name introduces a dedicated entity, the so-called \emph{authenticator} to the bus which is tasked with modifying a message's \ac{MAC} tag at line rate such that receivers can verify the source \emph{and} integrity of a message.
Meanwhile, the authenticator cannot generate valid tags itself.

We show how \name can be integrated into \ac{CAN} while offering full backwards compatibility and thus allowing for incremental deployment and operation alongside legacy \ac{CAN}.
\name could thus be deployed in the same manner as other extensions to \ac{CAN}, such as CAN-FD or the newly standardized CAN-XL.
Most importantly, existing receivers implementing AUTOSAR SecOC~\cite{autosar} can benefit from \name's source authentication without any modifications, as only senders need to support \name.

\vspace{2mm}
\textbf{Contributions.} To bring multicast source authentication to \ac{CAN}, we make the following contributions:
\vspace{1mm}

\begin{itemize}[itemsep=0pt,leftmargin=*,parsep=0pt,topsep=0pt]
    \item We propose \name, a multicast source authentication scheme specifically designed for bus communication with short authentication tags and no verification delay.
    \item We enable the calculation of \ac{MAC} tags by \name's authenticator at line rate by adapting the novel BP-MAC~\cite{2022_wagner_bpmac} scheme, whose security is based on AES.
    \item We design a \emph{reactive} bit flipping mechanism that enables dynamic overwriting of both recessive and dominant CAN bits after preemptively assessing their states.
    \item We prototypically integrate \name into CAN on a software-defined CAN controller~\cite{2019_cena_sdcc} and show backward-compatibility with real \acp{ECU} as well as reliability of over \SI{99.99}{\percent}~(similar to standard CAN) even for bus lengths of \SI{100}{\meter} and beyond.
\end{itemize}

\section{Background: CAN's Physical Layer}

The \ac{CAN} bus protocol is the de-facto standard in the automotive industry and mandatory in the EU for vehicle diagnostics and nowadays also finds applications in, \eg industrial automation~\cite{2017_li_industrial}.
To realize \name, we need to dive into some of the less well-known details of CAN's physical layer, which we outline in the following.

\textbf{Signaling.}
Physically, the CAN bus is based on two wires, CAN\_H (CAN high) and CAN\_L~(CAN low), to which every \ac{ECU} is connected~\cite{iso11898}.
At the physical layer, bits are encoded using a simple \ac{NRZ} scheme via different voltage levels specifying the following differential coding~\cite{iso11898-2}: 
%\vspace*{-0.1cm}
\begin{equation*}
\left\{ 
  \begin{array}{ c l }
   0 \textrm{  (dominant bit)} & \quad \textrm{if } CAN\_H - CAN\_L \geq 0.9\,V \\
   1 \textrm{  (recessive bit)} & \quad \textrm{if } CAN\_H - CAN\_L \leq 0.5\,V
  \end{array}
\right.
\label{eq:coding}
%\vspace*{-0.1cm}
\end{equation*}

%\mh{Move Figure 1 to this page?}

We exemplify the resulting differential signaling in \autoref{fig:can_frame}.
The transmitter only actively applies the voltage difference for dominant bits, while the recessive voltage levels resulting in a difference are passively created by resistors within each transceiver.
Due to this wired AND gate~(bus lines are recessive only when all transmitters transmit a recessive bit), dominant bits (0) always overwrite recessive bits (1)~\cite{iso11898}.
This property is utilized on the data link layer for arbitration and error handling.

\textbf{\Acl{ID} and Data Frame Format.}
Two \textit{data frame} formats with different \ac{ID} schemes for addressing are standardizes~\cite{iso11898}.
The base frame consists of an 11-bit \ac{ID} with up to  \SI{8}{bytes} of payload, three control flags~(3\,bits), the length field~(4\,bits), the \acs{CRC}~(16\,bits), and the \ac{ACK} field~(2\,bits) as shown in \autoref{fig:can_frame}.
The extended frame format offers a 29-bit \ac{ID}.
Both data frames are enclosed in a \ac{SOF} and an \ac{EOF} delimiter, which are signaled by one dominant and 7~recessive bits, respectively.

\textbf{Sampling and Synchronization.}
The fixed duration for which a single bit is present on the wire is called the \textit{nominal bit time} and depends on the used data rate.
Within this bit time, each signal is split into four time segments that consist of one or more \textit{time quanta} which are derived from the \ac{ECU}'s clock and usually are configurable by a programmable prescaler~\cite{iso11898}.
The first segment is exactly one time quanta long and is used to synchronize the different \acp{ECU}.
The second segment compensates for signal propagation and its length, in combination with the bit rate, are thus the main contributors to determining the maximal bus length.
The last two segments are chosen such that the bit sampling, happening between them, is located as close as possible to \SI{75}{\percent} of the nominal bit time~\cite{canopen}.
These two segments can also be elongated or shortened for resynchronization based on continuous monitoring of the edges of the voltage levels. 

\textbf{Bit Stuffing.}
With \ac{NRZ} coding, a certain number of consecutive bits of equal value leads to an absence of edges necessary for resynchronization, which is addressed by \textit{bit stuffing}.
To prevent that more than five identical bits appear in sequence in a bit stream, an additional bit with the inverse level is inserted (\textit{stuffed}) after five identical bits (cf.\ \autoref{fig:can_frame}).
%Bit stuffing is usually performed by the CAN controller and is transparent for the application processor executing the \acp{ECU} function.
Stuff bits are therefore transparently added and removed during transmission by the transmitting and receiving CAN controllers.

\section{State-of-the-Art on Securing CAN}

Historically, the CAN bus protocol offers no protection against cyberattacks.
Recent real-world demonstrations~\cite{2011_checkoway_comprehensive, 2017_nie_freefall, 2015_miller_remote, 2018_nie_over, 2020_wen_plug, 2015_foster_fast,2022_de_canflict , 2024_car_theft} have shown that this lack of security enables the remote compromise of entire vehicles.
A famous example takes full control of Tesla Model~S by remotely attacking the car's multimedia system and propagating from there through the CAN bus~\cite{2017_nie_freefall}.
Consequently, new protection mechanisms are needed to be prepared for the ever-increasing digitalization and interconnection of modern cars.

\subsection{Threat Model}

We consider a threat model that corresponds to that chosen in the majority of the recent offensive and defensive research on CAN security~\cite{2024_lotto_survey, 2019_bella_toucan,2016_nurnberger_vatican,2021_pese_s2can,2021_bhatia_evading,2011_checkoway_comprehensive, 2015_foster_fast,2015_miller_remote,2017_nie_freefall,2018_nie_over,2014_woo_practical}:
We assume that an attacker has either physical access to the bus to implant a malicious \ac{ECU} or has compromised an existing \ac{ECU}, \eg through Bluetooth or other connectivity.
The attacker then intends to inject messages beyond those message types needed for the functionality of the compromised \ac{ECU}~(an implanted malicious \ac{ECU} is not supposed to send any messages), \ie masquerading as another \ac{ECU}, to highjack the car.

\begin{table*}[t]% Try here, and then top
    
    \centering \footnotesize

    \newcolumntype{S}[0]{>{\Tstrut}m{4.1cm}<{\Bstrut}}
    \newcolumntype{Z}[0]{>{\Tstrut\centering}m{1.2cm}<{\Bstrut}}
    \newcolumntype{Y}[0]{>{\Tstrut\centering}m{6.45mm}<{\Bstrut}}

    \setlength{\tabcolsep}{3.75pt}
    
    \begin{tabularx}{\textwidth}{ cSZ YYYY c YYYY c YYY}
            
            % HEADER
            &&&\multicolumn{4}{l}{\textbf{Limitation}}&&\multicolumn{4}{c}{\textbf{Vulnerability}}&&\multicolumn{3}{c}{\textbf{Overhead}}\tabularnewline
            \cline{4-7}
            \cline{9-12}
            \cline{14-16}
            \noalign{\smallskip} 
            &\textbf{Protocol}
            & \textbf{Year}
            & \nobroadcast
            & \scalability
            & \desync
            & \verificationdelay
            &
            & \disconnect
            & \masquerade
            & \replay 
            & \detectiondelay 
            &
            & \storage 
            & \communication 
            & \delay \tabularnewline\midrule
            \multicolumn{2}{l}{\parbox{4.85cm}{IDSs~\cite{2017_cho_viden,2018_choi_voltageids, 2019_foruhandeh_simple, 2020_kneib_easi, 2016_cho_fingerprinting, 2018_wang_delay, 2019_young_automotive,2023_schell_sparta, 2023_shin_ridas, 2022_rogers_detecting, 2023_nichelini_canova, 2018_kneib_scission, 2023_longari_candito, 2019_longari_copycan, 2011_muter_entropy,2016_song_intrusion}}} 
            &2016- 2023&
            - & - & - & - && \full & \part & - & \part && - & - & -
            \tabularnewline\midrule
            \arrayrulecolor{lightgray}

            \parbox[t]{6mm}{\multirow{4}{*}[5pt]{\rotatebox[origin=c]{90}{\parbox{2cm}{\centering Covert\\Channels}}}}
            &CANTO~\cite{2020_groza_canto} &2020& 
            - & - & - & - && - & \full & - & - && - & - &\full\tabularnewline\cline{2-16}
            
            &Watermarking~\cite{2022_michaels_watermarking} &2022&  
            - & - & - & - && - & \full & - & - && - & - &-\tabularnewline\cline{2-16}
            
            &ZBCAN~\cite{2023_serag_zbcan} &2023& 
            - & - & - & \part && \full & - & \part & - && - & - & \part
            \tabularnewline\cline{2-16}
            \arrayrulecolor{black}
            
            &CAN-MM\cite{2024_oberti_canmm} &2024&
            - & - & - & - && - & \full & - & - && - & - & -
            \tabularnewline\midrule

            \parbox[t]{3mm}{\multirow{12}{*}[-16mm]{\rotatebox[origin=c]{90}{Cryptographic Approaches}}}
            &AUTOSAR SecOC~\cite{autosar} &2020& 
            - & - & - & - && - & \full & - & - && - & \part & - \tabularnewline\cline{2-16}
            \arrayrulecolor{lightgray}

            &CANAuth~\cite{2011_van_canauth} &2011& 
            - & - & - & - && - & \full & - & - && \full & \part & - \tabularnewline\cline{2-16}

            &Car2X~\cite{2011_schweppe_car2x} &2011& 
            - & - & - & \full && - & - & \full & - && \part & \full & - \tabularnewline\cline{2-16}

            &LiBrA-CAN~\cite{2012_groza_libra} &2012& 
            - & \full & - & \full && - & \part & - & - && - & \full & - \tabularnewline\cline{2-16}

            &LinAuth~\cite{2012_lin_cyber} &2012&
            - & \full & - & - && - & - & - & - && \full & \full & - \tabularnewline\cline{2-16}

            &LCAP~\cite{2012_hazem_lcap} &2012&
            - & - & - & - && - & - & \full & - && \full & \full & - 
            \tabularnewline\cline{2-16}

            &CaCAN~\cite{2014_kurachi_cacan}  &2014&
            - & - & \full & - && \full & \part & - & - && - & \part & - 
            \tabularnewline\cline{2-16}

            & Woo-Auth~\cite{2014_woo_practical} &2014&
            - & - & \full & \part && - & \full & - & - && \part & \part & - 
            \tabularnewline\cline{2-16}

            &VeCure~\cite{2014_wang_vecure} &2014& 
            - & - & \full & \part && - & \full & - & - && \full & \full & - 
            \tabularnewline\cline{2-16}

            &LeiA~\cite{2016_radu_leia} &2016&
            - & - & - & \full && - & \full & - & - && \full & \full & - 
            \tabularnewline\cline{2-16}

            &vatiCAN~\cite{2016_nurnberger_vatican} &2016&
            - & - & \part & \full && - & \full & \part & - && \part & \part & - 
            \tabularnewline\cline{2-16}

            &VulCAN~\cite{2017_vanbulck_vulcan} &2017& 
            - & - & - & \full && - & \full & - & - && \full & \full & - 
            \tabularnewline\cline{2-16}

            &TOUCAN~\cite{2019_bella_toucan} &2019&
            - & - & - & - && - & \full & - & - && - & \part & -  
            \tabularnewline\cline{2-16}

            &\Tstrut\Bstrut LEAP~\cite{2019_lu_leap} &2019&
            \full & - & - & - && - & - & - & - && \full & \part & - 
            \tabularnewline\cline{2-16}

            &CAN-TORO~\cite{2020_groza_highly} &2020&
            - & - & - & - && - & \full & \part & - && \full & - & - 
            \tabularnewline\cline{2-16}

            &MAuth-CAN~\cite{2020_jo_mauth} &2020&
            - & - & \part & \full &&  \full & \part & - & - && - & \part & - 
            \tabularnewline\cline{2-16}
            
            &AuthentiCAN~\cite{2020_marasco_authentican} &2020&
            \full & - & - & - && - & - & - & - && \full & \full & - 
            \tabularnewline\cline{2-16}

            \arrayrulecolor{black}
            &S2-CAN~\cite{2021_pese_s2can} &2021& 
            - & - & - & - && - & \full & - & - && - & \part & - 
            \tabularnewline\cline{2-16}

            \arrayrulecolor{black}
            & \name & 2025 &
            - & - & - & - && - & - & - & - && - & \part & -
            \tabularnewline\midrule
 
            \multicolumn{13}{l}{ \parbox{.78\textwidth}{
                \scriptsize
                Limitation:\hfill\nobroadcast~no broadcasting support\quad\scalability~limited scalability\quad\desync~no resynchronization\quad\verificationdelay~delayed verification  \newline
                Vulnerability:\hfill\disconnect~covertly disconnected device\quad \masquerade~masquerading by other ECU\quad\replay~frame interception\quad \detectiondelay~delayed alarms \newline
                Overhead:\hfill\storage~storage overhead\quad\communication~communication overhead\quad\delay~delayed transmission
            }} & 
            \multicolumn{3}{r}{\raisebox{2ex}[0ex][0ex]{\full\,full \quad\part\,partial}}

    \end{tabularx}

    \caption{
        Current proposals to protect CAN traffic expose weaknesses that make them either not deployable in cars (\eg scalability limitations) or offer attack vectors to malicious actors (\eg no protection against masquerading).
    }
    \label{tab:related-work}

\end{table*}

\subsection{Related Work}
\label{ssec:rw}

%In recent years, a broad range of research has been conducted to .
Research on protecting CAN against malicious ECUs can be grouped into three categories:
\emph{intrusion detection systems~(IDSs)}, \emph{covert channels}, and \emph{cryptographic approaches}.
These approaches make certain trade-offs to achieve alleged security, which result in certain  limitations, vulnerabilities, or overhead.
We classify the current state-of-the-art in protecting CAN according to these drawbacks as introduced in the following.

\vspace{2mm}
\noindent\textbf{Limitations} 
\vspace{1mm}

\begin{itemize}
    \setlength\itemsep{.3em}

    \item[\nobroadcast] Only point-to-point communication is protected, whereas broadcast communication, \ie CAN messages meant for multiple receivers, are not protected.
    \item[\scalability] Only broadcast messages with few (\eg $<$\,5) receivers are protected, where the level of protection exponentially reduces with the number of receivers.
    \item[\desync] Definitively lost CAN frames lead to desynchronizations that cannot be recovered from.
    \item[\verificationdelay] The protocol deterministically delays message authenticity verification by buffering messages at either the sender or receiver. 

\end{itemize}

\vspace{2mm}
\noindent\textbf{Vulnerabilities}
\vspace{1mm}

\begin{itemize}
    \setlength\itemsep{.3em}

    \item[\disconnect] The physical disconnection of a single ECU~(\eg an IDS) covertly disables all protection mechanisms in a way that is unnoticeable by any other ECU.
    \item[\masquerade] An attacker that compromises a single CAN ECU can subsequently masquerade as other ECUs, \eg because all ECUs share a group key. Needing to compromise a dedicated security node (\eg IDSs) is considered more secure, as those offer no outside connectivity and should be temper-resilient.
    \item[\replay] An attacker can intercept a frame and use the information gained to impersonate another device, \eg by replaying the frame at a later time. Some approaches may only enable the injection of malicious frames shortly after the original intercepted frame.
    \item[\detectiondelay] The detection of malicious messages only happens retroactively after messages have been processed and potentially caused significant harm.

\end{itemize}

\vspace{2mm}
\noindent\textbf{Overhead}
\vspace{1mm}

\begin{itemize}
    \setlength\itemsep{.3em}

    \item[\storage] The scheme needs excessive storage overhead, \eg to store a secret key for each other ECU.
    \item[\communication] The scheme requires some additional communication overhead, either by reserving some space in each frame or even transmitting additional CAN frames.
    \item[\delay] The transmission of CAN frames is delayed or prioritization is not fully adhered to.

\end{itemize}
\hfill\vspace*{-2mm}

We consolidate our analysis of the state-of-the-art on protecting CAN in \autoref{tab:related-work}.
Our classification is partially based on a recent survey by Lotto~\etal~\cite{2024_lotto_survey} on weaknesses in CAN authentication protocols.
For the five proposals classified as \textit{secure} by Lotto~\etal and seven proposals not considered in their survey, we present a detailed analysis in \hyperref[app:cryptanalysis]{Appendix~\ref{app:cryptanalysis}} to substantiate our classification.
In the following, we give an overview of the three categories of CAN protection mechanisms.

%Intrusion detection systems~\cite{2016_cho_fingerprinting, 2018_choi_voltageids, 2019_young_automotive, 2018_kneib_scission,2020_kneib_easi,2023_schell_sparta, 2017_cho_viden,2019_foruhandeh_simple,2019_longari_copycan,2022_rogers_detecting, 2023_shin_ridas, 2023_longari_candito, 2023_nichelini_canova} 
\textit{\acp{IDS}} add a %designated
dedicated device to the network that monitors the physical characteristics and behavior of all \acp{ECU} to detect imposters.
This monitoring can be based on voltage levels~\cite{2017_cho_viden,2018_choi_voltageids, 2019_foruhandeh_simple, 2020_kneib_easi}, message timings~\cite{2016_cho_fingerprinting, 2018_wang_delay, 2019_young_automotive,2023_schell_sparta, 2023_shin_ridas, 2022_rogers_detecting, 2023_nichelini_canova,2016_song_intrusion}, signal characteristics~\cite{2018_kneib_scission, 2023_longari_candito, 2023_nichelini_canova}, or device behavior~\cite{2019_longari_copycan, 2011_muter_entropy}.
All methods have in common that the IDS device can be disrupted to covertly disable security without this being noticed \disconnect.
Moreover, IDSs often only detect malicious message flows rather than achieving single-message detection, such that timely reactions, \eg discarding messages, are hardly possible \detectiondelay~\cite{2023_serag_zbcan}.
Finally, IDSs have no perfect detection performance, \ie malicious messages may not get recognized~(false negative) or genuine traffic falsely gets flagged~(false positive). 
Even %extremely 
low false positive rates can have detrimental effects if acted upon, given the amount of genuine CAN messages being constantly sent.
On top, recent research shows how sophisticated attacks can also deliberately evade IDSs %'s detection mechanisms 
in CAN~\cite{2021_bhatia_evading,2018_sagong_cloaking}.
Overall, we conclude that IDS may be useful, but should not be the only line of defense due to their limitations.

A second class of CAN security approaches relies on \textit{covert channels} to achieve message authentication.
These covert channels can be created through a high-frequency signal interlaced with regular messages~\cite{2022_michaels_watermarking,2024_oberti_canmm, 2020_groza_canto}, which however requires dedicated transceiver hardware for decoding and still does not provide source authentication~\masquerade.
ZBCAN~\cite{2023_serag_zbcan}, on the other hand, proposes %to authenticate senders through 
a unique inter-frame spacings for each sender, only known to the sender and a central authority. %, for authentication.
In case of anomalies, %inconsistencies, 
the central authority jams the suspicious frame.
However, in ZBCAN, it is neither detectable that the central authority is covertly disconnected~\disconnect~nor is the protocol secure against bit modification attacks~\cite{2023_yousik_bit_modification} as it only verifies that the sender intended to send at a given time but not the content of the message~\replay.
Moreover, ZBCAN modifies CAN's prioritization scheme which can lead to a significant delay of up to \SI{25}{\milli\second}~\verificationdelay. % upon message reception.

% cite secure work from survey
Finally, \textit{cryptographic approaches} %~\cite{autosar, 2019_bella_toucan, 2016_nurnberger_vatican, 2021_pese_s2can, 2016_radu_leia, 2012_lin_cyber, 2012_groza_libra} 
rely on the integration of integrity-protecting tags, usually \acp*{MAC}, into data frames.
These tags can be transmitted \eg as a portion of the payload~\cite{autosar}, as a substitution for the \ac{CRC} checksum~\cite{2014_woo_practical}, as part of the CAN ID~\cite{2020_groza_highly}, or in an additional frame~\cite{2016_nurnberger_vatican}.
A prominent example is the AUTOSAR SecOC standard~\cite{autosar} that uses part of the payload, \eg \SI{28}{\bit}, for the transmission of integrity protection.
Like AUTOSAR SecOC, most of these approaches rely on \emph{group keys} to compute integrity tags.
The resulting lack of source authentication means that compromised \acp{ECU} are not restricted from simply impersonating other devices~\masquerade.
The few exceptions split the limited space for integrity protection among all receivers~\scalability~\cite{2012_groza_libra,2012_lin_cyber}, do not support broadcast communication~\nobroadcast~\cite{2019_lu_leap, 2020_marasco_authentican}, rely on a covertly disconnectable authenticator~\disconnect~\cite{2014_kurachi_cacan,2020_jo_mauth}, or are vulnerable to frame interceptions~\replay~\cite{2011_schweppe_car2x,2012_hazem_lcap}.

Concluding, we observe that all current proposal to protect CAN suffer from drawbacks that either make them unsuitable for in-vehicle communication or make them vulnerable to attacks.
With \name, we thus aim to design a protocol that suffers none of these drawbacks.

\section{Source Authentication with \name}
\label{sec:caiba}

For traditional point-to-point communication, source authentication and integrity protection is realized by appending an integrity-protecting authentication tag (\ie a MAC) to each message.
This tag is computed with the help of a secret key shared by sender and receiver, and verified by the receiver upon reception of a message.
However, in a broadcasting domain like CAN, such authentication tags do not provide source authentication:
A receiver cannot differentiate if a message stems from the alleged sender or if it has been spoofed by another member in the group of receivers who has access to the secret key.
Hence, a compromised device could masquerade as one or multiple other devices.
To protect the CAN bus against masquerading attacks, an effective source authentication protocol for multicast communication is thus required.
In the following, we first investigate existing proposals to achieve source authentication and argue why they are not suitable for CAN.
Afterward, we propose our novel scheme, \name, that addresses limitations of prior work.

\subsection{The Current State-of-the-Art}
\label{sec:caiba:stateoftheart}

Source authentication for multicast communication relies on asymmetry %of secrets %\tz{of secret information/key material... Einfach nur asymmetrie kann ja alles mögliche sein. Nähme natürlich viel Platz in dieser Section, irgendwie abkürzen?} 
between senders and receivers or across receivers~\cite{2004_challal_taxonomy}.
The most widespread form of asymmetry are digital signatures, where receivers only have keys to verify messages, but not authenticate them.
The main concern for this type of \emph{cryptographic asymmetry} is the size of signatures~(upwards of \SI{32}{bytes}) that cannot fit into CAN frames.
The TESLA protocol~\cite{2000_perrig_tesla,2001_perrig_tesla2}, on the other hand, relies on \emph{time asymmetry} for multicast source authentication.
Here, keys are revealed after they have expired and linked to the sender~(\eg via a hash-chain), such that receivers can verify the authenticity of buffered messages retroactively.
Time asymmetry does however introduce significant delays and some additional bandwidth overhead~(\eg for key revelation) which are not acceptable in in-vehicular communication.
The final type of multicast source authentication schemes relies on \emph{information asymmetry}, where receivers can only verify parts of the multiplexed source-verifying information~\cite{1999_canetti_multicast, 2001_perrig_biba}.
Thus, each receiver has reduced certainty of a message's authenticity, with the benefit that no single receiver can generate an authentication tag that is accepted by all receivers.
This, however comes at the cost of longer authentication data (compared to simple MAC schemes), which grows with the number of receivers.
To conclude, current multicast source authentication schemes are inapplicable to CAN: they either incur verification delay or require excessive message overhead~(in CAN, payload and authentication tag must fit into \SI{8}{bytes}).

\subsection{General Idea of \name}
\label{sec:caiba:idea}

\begin{figure}
    \centering
    \includegraphics[width=0.97\columnwidth]{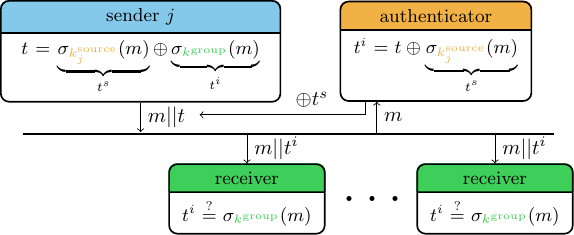}
    \caption{
        \name employs an integrity-protecting tag $t^{i}$ and a source-authenticating tag $t^{s}$ that are computed by the MAC function $\sigma$ using the source and group keys, \ksource and \kgroup, and 
        are XORed %together 
        by the sender to protect a message $m$.
        During transmission, the authenticator computes $t^{s}$ and overwrites the message such that only $t^{i}$ remains.
        The integrity-protecting tag $t^{i}$ that is read by the receivers can then be verified with a conventional group key without knowledge of the source key.
    }
    \label{fig:concept}
\end{figure}

To enable multicast source authentication for CAN, we devise the novel \name scheme.
To understand the idea behind \name, consider the following hypothetic scenario.
A CAN message is protected by two authentication tags, the \emph{integrity-protecting tag} $t^{i}$ and the \emph{source-authenticating tag} $t^{s}$.
These tags are computed with traditional \ac*{MAC} schemes, where the tag verification corresponds to the re-computation of the tag based on the received data and comparison to the received tag.
The integrity-protecting tag $t^{i}$ is generated using a group key \kgroup and verified by each receiver.
Like intrusion detection systems and other proposals to retrofit integrity protection into CAN~\cite{2012_groza_libra, 2014_kurachi_cacan, 2023_serag_zbcan}, \name relies on one or multiple dedicated security nodes, the authenticator(s).
The source-authenticating tag $t^{s}$ is only verifiable by this authenticator that shares a symmetric key \ksource with each sender $j$, but does not know the group key.
In this hypothetic scenario, the authenticator also controls a virtual side-channel to securely notify each receiver whether or not the source of a message has been correctly verified.

Obviously, this basic construction occupies additional space~(to send two tags instead of one) and time~(to wait for the confirmation of the source-authenticating tag $t^{s}$ verification to receivers).
However, in \name, we only transmit a single tag and use an implicit side-channel as shown in \autoref{fig:concept}:
The actual transmitted tag $t$ is the aggregation of both tags, \ie $t = t^{i} \oplus t^{s}$.\footnote{\,The aggregation with XOR is provably secure and $t$ can be verified by recomputing $t^{i}$ and $t^{s}$ individually~\cite{1995_bellare_xormac}.}
The authenticator no longer verifies $t^{s}$ but only recomputes $t^{s}$ based on the received payload and the alleged sender.
The authenticator then XORs the tag $t$ with $t^{s}$ as the bits are transmitted over the bus.
When the ordinary receivers in CAN sample the bus, they will then read and verify $t^{i}$.
If, and only if, neither the data nor the tag has been manipulated beyond the authenticator's actions, the receiver's integrity verification can be successful.
In this way, no explicit side-channel communication is needed and only the space of a single authentication tag is occupied in the CAN payload.

\subsection{Requirements to Deploy \name}
\label{sec:caiba:requirements}

Since \name introduces no verification delay or message overhead compared to an ordinary group key based authentication, our scheme is especially attractive for the CAN bus.
However, as other solutions for multicast source authentication, \name can only be applied if certain requirements are fulfilled.
Specifically, the deployment scenario must fulfill certain requirements.\vspace{1mm}

\textbf{No Direct Communication.}
Entities must not be able to directly communicate with each other without the authenticator also overhearing these transmissions.
Otherwise, a malicious group member could only append the integrity-protecting tag $t^{i}$ and a receiver would have no way of knowing whether the message really stems from the alleged transmitter.
Communication buses like CAN fulfill this requirement by design.\vspace{1mm} % or %just 
%any other potential.

\textbf{Ability to Overwrite Messages.}
The authenticator must be able to reliably %and quickly 
overwrite messages at line rate.
We later show how this is possible for CAN~(\cf~\autoref{sec:integration:authenticator}). % a custom MAC and hardware components.
%But it is not a given that such overwriting is always possible based on a protocol's specification.\vspace{1mm}
For other communication protocols, it still has to be investigated how to unlock such capabilities.\vspace{1mm} 

\textbf{No Authenticator Collusion.}
There must be no collusion between the authenticator and any other entity in \name.
Otherwise, the entity could transmit a message without a source-authenticating tag and the colluding authenticator would simply ignore it. % not alter it.
Practical demonstrations have shown how to compromise individual CAN ECUs due to external connectivity.
However, additionally compromising an authenticator with no external connectivity that can rely on temper-resilient hardware would require a significantly more advanced attack.

\subsection{Security Discussion}
\label{sec:caiba:security}

As we proof in \hyperref[app:proof]{Appendix~\ref{app:proof}}, \name provides the same security as a traditional \ac*{MAC} %tag 
of the given length used to protect an end-to-end connection between sender and receiver.
Considering an exemplary \SI{24}{bit} tag, an attacker thus has a mere $1/2^{24}$ ($\sim$1 in 17\,million) chance to guess a valid tag.
This security level is achieved because the aggregation of the source-authenticating tag and the integrity-protecting tag achieves the same security level as the individual tags~\cite{2008_katz_aggregate}.
The difference is that the aggregated tag can only be verified by combining the knowledge to verify both tags individually.
As only the sender knows the keys for both tags, a valid tag authenticates the source of a message to each receiver.

One security risk for \name is that an attacker physically disconnects the authenticator from the network.
With other security solutions, \eg IDSs in CAN, such attacks are often not detectable, and the attacker would gain complete control over the bus.
In contrast, \name quickly detects that the authenticator is disconnected and can take corrective actions before significant damage can be caused.
Ideally, redundant authenticators are available, which can quickly take over upon request by the sender that noticed an inactive authenticator.
Alternatively, operators can be notified about the inoperational authenticator and respond similarly as to an alarm by an IDS.
Luckily, an authenticator's physical disconnection most likely occurs while the car is stationary, such that a car could be prevented from moving, at least until the driver is notified about the risk.
If the authenticator nonetheless disconnects while the car is moving, \name could fall back to insecure CAN, advise the driver to stop the car, and potentially the speed and acceleration of the car could be limited.

Finally, an attacker could compromise the authenticator directly, \eg through a supply chain attack.
However, even a compromised authenticator cannot spoof messages as it cannot compute integrity-protecting tags.
To successfully spoof a message, an attacker needs to additionally compromise a genuine receiver within the corresponding multicast group.
Thus, \name overall provides strong protection against masquerading attacks and enables the detection of a disconnected authenticator.

\section{Integrating \name into the CAN Bus}
\label{sec:integration}

After presenting the idea of \name in a general and abstract fashion, we now discuss the technical details of integrating \name into CAN. 
Here, we are faced with two main challenges.
First, the authenticator must be able to quickly compute the source-authenticating tag $t^{s}$ as the first bit of the overwritten tag follows immediately after the last bit of the protected payload.
Secondly, the authenticator must be able to dynamically and precisely flip individual bits.
In the following, we will successively introduce the design of all entities,
%three communication components, 
\ie the transmitter, the authenticator, and the receiver.
Before, we shortly address the general challenges of deploying \ac{CAN} extensions such as \name.
Concerning key distribution, % distribution of key material, 
we assume that each \ac{ECU} and the authenticator are initially configured with exactly those keys that they require, \ie each \acp{ECU} shares a unique key with the authenticator, and all \acp{ECU} know relevant group keys\footnote{\,There maybe only exists a single group key for the entire bus.}.

\subsection{Deployment Considerations}
\label{sec:integration:deployment}

CAN extensions such as CAN-FD for higher bandwidth have in the past been successfully deployed by ensuring interoperability with legacy CAN devices.
Thus, while \acp{ECU} do not necessarily need to be fully compliant with the existing \ac{CAN} standard, any changes must by compatible with the existing standard, such that \ac{CAN} \acp{ECU} and \name \acp{ECU} can operate on the same network.
Only then can an incremental deployment of \name be possible, as the sudden adoption of a new standard by all actors in, \eg a car manufacturing supply chain, is not realistic.
The easiest solution to achieve such interoperability is if all changes are restricted to the inner operation of an individual \ac{ECU}, while the signals written to the bus are fully compliant to the \ac{CAN} standard at all times.

Regarding the actual integration and commercialization of Caiba into \eg cars, ECU and car manufacturers do not have to change much in their operations.
Nowadays, CAN controllers are mostly integrated as \ac{SIP} cores (e.g., \cite{amd,nxp})
Once an SIP core implements Caiba, ECU manufacturers can integrate them into newly manufactured chips.
Then, car manufactures need to add an authenticator to their bus and configure it for \name-supporting \acp{ECU} to talk securely if all receivers interpreting a specific frame implement the AUTOSAR SecOC standard.

\subsection{Transmitter Design}
\label{sec:integration:transmitter}

Transmitters of CAN frames only require minor changes to support \name.
These changes, however, require little additional space on the die of the ECU chip.

The main change between a transmitter supporting the AUTOSAR SecOC standard and a transmitter supporting \name is the computation of two authentication tags instead of one over the concatenated CAN ID and payload.
For the source-authenticating tag $t^{s}$, we rely on BP-MAC, a novel MAC scheme optimized for short messages of only a few bytes~\cite{2022_wagner_bpmac}.
%It realizes its unique performance by bit-wise precomputing the tag.
%It is precisely this design that we can take advantage of to realize tag computation at line-rate for the authenticator.
The integrity-protecting tag $t^{i}$ can be computed by any suitable MAC scheme, which also allows the deployment of \name without any modifications to the receivers, as they only receive and verify this tag.
Once both tags are computed, the transmitter simply aggregates them with XOR for the transmitted tag $t$ and integrates it into the payload of the CAN frame.
We use \SI{24}{bits} reserved in the payload for the tag and keep track of a counter for replay protection by appending its 4 least-significant bits in each frame.
The usage of a \SI{24}{bits} MAC with the 4 least-significant counter bits corresponds to \code{SecOC Profile 3 (JASPAR)}~\cite{autosar}.
%By using \SI{24}{bits} for the tag, the probability of an attacker randomly guessing a valid tag is 1 in $\sim$17\,million.

Furthermore, the transmitter must be adapted to support tag overwriting by the authenticator. 
On the one hand, the CRC checksum and the placement of stuff bits in the final received frame must be ensured.
Therefore, the sender computes the checksum and the stuff bits placement based on the expected final frame, after the authenticator's modification, and not based on the transmitted frame.
The authenticator knows when to expect stuff bits and skips over them.
For the particular CAN controller used for our implementation, this requirement was naturally fulfilled:
The controller listens to the bus while transmitting its message to detect higher-priority transmissions and uses this received data for bit stuffing and \ac{CRC} computation.
To the best of our knowledge, this behavior is however not required by the standard.

On the other hand, the transceiver must ignore overwritten signals during the transmission phase of the tag.
Otherwise, the transmitted would quickly switch to the bus-off state and virtually disconnect itself from the bus.
Therefore, we change bit monitoring to listen to the expected bits after an authenticator overwrote the bits.
This change does not lead to security or reliability problems, since every unauthorized modification would result in an invalid tag on the receivers' side.
If errors only increase during the tag transmission than an \ac{ECU} can conclude that the authenticator is faulty or disabled and react accordingly.
This reaction could be the fallback to unsecured \ac{CAN} after alerting the network or the switch to on a backup authenticator.

While \name requires some modifications to CAN transmitters, it is important to note that the communication on the bus still remains \ac{CAN} compliant and interoperable.
Therefore, \name can coexist with legacy CAN transmitters as long as the authenticator knows which communications are \name-protected and which are not.

\subsection{Authenticator Design}
\label{sec:integration:authenticator}

The authenticator continuously monitors the bus and notices when the transmitter starts writing to the bus.
%If it detects the transmission of a message, 
Then, it has to \emph{(i)}~identify the transmitter, \emph{(ii)}~compute the source-authenticating tag $t^{s}$ over the CAN ID and the first part of the payload as well as \emph{(iii)}~overwrite the second part of the payload based on that computed tag.
All of this processing needs to happen concurrently with the ongoing transmission of the CAN frame.
In the following, we discuss how these three steps are realized in \name.

\subsubsection{Source Key Identification}

The authenticator first identifies the alleged source of a frame and whether it supports \name\footnote{\,We assume that the bus is configured according to AUTOSAR SecOC, \ie space for authentication tags is reserved and nodes not capable of performing the integrity verification ignore the tag. Alternatively, all ECUs must track for which CAN IDs integrity protection is enabled.}.
While CAN uses message identifiers that should be associated with a unique sender, often employed standards %such as the J1939 protocol
ensure an easy coupling without a large lookup table~\cite{2024_lotto_survey}.
Thus, the message ID can be used to identify the unique transmitter of a message.
It is crucial to securely configure which ECU supports \name during vehicle assembly to thwart downgrade attacks.
After the sender has been decoded by the authenticator, the transmitter and thus the relevant key \ksource is available.

\subsubsection{Fast In-Line MAC Computation}

When a \name transmitter is identified, the authenticator computes the source-authenticating tag $t^{s}$ based on
%The input for this computation is
the source authentication key \ksource, the protected payload, and a nonce (a \SI{8}{bytes} long counter which is synchronized based on the 4 least-significant bits transmitted with each frame).
Considering CAN's maximum supported data rate of \SI{1}{Mbit/s}, the first bit of the tag follows within \SI{1}{\micro\second} after the last data bit.
Actually, the authenticator needs to compute $t^{i}$ in a fraction of this time, as it needs time to overwrite this first tag bit before it is read by the receiver.

To achieve these speeds while still relying on sound cryptography, \name uses the BP-MAC~\cite{2022_wagner_bpmac} scheme with its unique performance through bit-wise precomputing tags and enhances it by a custom online computation algorithm.
For now, it is only important that the computation of the source-authenticating tag $t^{s}$ can be achieved with a single XOR operation per read bit, which is fast enough to be performed even in a fraction of the time between two CAN signals.
Meanwhile, BP-MAC relies on a Carter-Wegman construction such that it offers the same guarantees as the underlying MAC scheme, \eg \code{HMAC\_SHA256}.
The authenticator is thus fast enough to know the source-authenticating tag $t^{s}$ that it must XOR with the transmitted tag $t$ as soon as its first bit is written to the bus.
The details of the BP-MAC %based 
tag computation in \name are covered in \autoref{sec:mac}.

\subsubsection{Overwriting CAN Frames}
%\tz{Wie schon geschrieben, ich finde diesen Absatz sehr gedoppelt mit 7, brauchen wir hier so ein detailliertes Niveau?}
Once the authenticator has read the source-authenticating tag~$t^{s}$, it must XOR this value with the transmitted tag, \ie whenever a bit in $t^{s}$ is set, the signal on the bus must be inverted. %\cut{, while also considering potential stuff bits}.
%Here, the authenticator also must consider potential stuff bits written by the sender and adapt the overwriting accordingly.
For the signal-flipping procedure, the authenticator first needs to receive the transmitted signal.
%By sampling early in second bit quantum, the authenticator can both read the originally transmitted signal but is still able to modify the bit early enough that other receivers which are configured to sample at the regular sample points receive the modified bit.
By sampling early in the nominal bit time, the authenticator can read the original signal while still having time to react.
This early sampling is similar to the process how CAN-FD achieves higher bitrates than CAN.
It is possible because of relatively large tolerances for rise and propagation times in the CAN standard, which are designed for the worst case where the transmitter and receiver are located as far apart as possible.
We suggest that the authenticator is placed centrally for bus length close to the maximum supported for a given bitrate, such that its distance from the transmitter is at most half of what is compensated for by CAN (even less if multiple authenticators are used).
If the bit on the bus is recessive, the authenticator can simply overwrite it with a dominant bit.
However, the opposite situation is not trivial, as dominant bits %should -- as the name suggests -- 
always overwrite recessive ones.
%To overwrite dominant bits with recessive ones, 
For this purpose, we take advantage of the limited output current of CAN transceivers by simply connecting an additional \emph{inverted} transceiver in our \ac{RBF} approach, which is explained in more detail in \autoref{sec:bitflip}.
%Whenever this additional transceiver shall overwrite a dominant with a recessive bit, it electrically zeroes out the written bit of the original transmitter, resulting in the recessive state of the bus lines.
%TZ: Was:
%Once the authenticator knows the source-authentication $t^{s}$, it must XOR this value with the transmitted tag, \ie whenever a bit in $t^{s}$ is set, the signal on the bus must be flipped.
%Here, the authenticator also must consider potential stuff bits written by the sender and delay \jb{hier reicht doch ein delay nicht immer, oder? besser "adapt"?} overwriting accordingly.
%For the signal flipping procedure, the authenticator first must know the transmitted signal.
%Here, we use an early sampling in the second bit quantum which is possible because the authenticator is placed centrally on the bus \jb{war das eine requirement? Sonst besser "if" statt "because".} such that signal propagation delays are lower.
%If the bit on the bus is recessive, the authenticator can simply overwrite it with a dominant bit, a process that is already used for CAN's arbitration mechanisms~(\cf~\autoref{sec:bg:dll:arbitration}).
%To overwrite dominant bits with recessive ones, we erase the voltage difference with a second CAN transceiver at the authenticator.
%This overwriting procedure is explained in more detail in Section~\ref{sec:bitflip}.
With \ac{RBF}, the authenticator can ultimately XOR the transmitted tag $t$ with the computed tag $t^{s}$, such that receivers sense the integrity-protecting tag $t^{i}$ when sampling the bus at standard sampling points.

\subsection{No Need to Adapt Receivers}
\label{sec:integration:receiver}

The design of \name requires no software or hardware changes at the receiver if the AUTOSAR SecOC standard is already supported since MACs are already implemented.
The modifications of \name only overlay the traditionally transmitted integrity-protecting tag $t^i$ with the source-authenticating tag~$t^s$.
However, this change is transparent for an \ac{ECU} which samples the bus as defined by the CANopen standard~\cite{canopen}.
%For no interference to happen, \name does not even need to reduce the maximally supported bus length~(see Section~\ref{sec:eval:length} for more details).
Thus, receivers do not need to be changed or replaced to support \name which enables a smooth and iterative deployment of \name.

Once a car manufacturer integrates an authenticator module, all CAN transmitters can decide to employ \name.
Each sending \acp{ECU} employing \name needs to share a secret key \ksource with the authenticator, which will, in most cases, be statically configured by the manufacturer.
Afterward, the authenticator starts overwriting the authentication tag according to \name's design.
A receiver can then process CAN frames exactly as before.
If a message is tampered with, either through manipulations during transmission or by being sent from an unauthentic source, the verification by the receiver fails and it processes this anomaly accordingly.

\subsection{Error Handling and Recovery}
\label{sec:integration:recovery}

Failed CAN transmissions could result from de-synchronized nonce counters of the sender, authenticator, and receiver.
If this were the case, no future message would be verifiable.
However, if CAN transmissions are not received by all \acp{ECU}, \eg due to an error in the CRC checksum at a single \ac{ECU}, this is announced with a distinct error frame.
In that case, the authenticator resets its counter and the sender retransmits the frame with the original MAC.
The receiver would not have processed the MAC as the error frame interrupts the transmission, such that no action is required.

Due to the transmission of the four least-significant bits of the nonce in each frame, it is unlikely that a node gets out of sync.
To get to such a state, one \ac{ECU} must miss an error frame or a frame must be discarded due to multiple failed retransmissions, at least 16 times in a row from the same origin.
If such an unlikely scenario were to occur, it most likely stems from a malfunctioning \ac{ECU} or an active denial of service attack, both of which \name cannot protect against.
Still, to ensure the best possible resilience, \name implements a distinct recovery mechanism for such cases.
If a receiving node fails to verify the tag of five consecutive frames, it requests a counter reset through a specific CAN ID.

Once this request is received by the sending ECU, it updates the two counters for the integrity-protecting and source-authenticating tags.
Therefore, the value in the $n$ most-significant bytes of the counter is incremented, where $n$ represents the number of payload bytes in a CAIBA frame.
The less-significant bytes are meanwhile reset to zero.
The sender then first sends the most significant bytes of the counter for the source-authenticating tags to the authenticator. 
This CAN frame is protected by a regular MAC (sender and authenticator share a secret key) and the frame must only be authenticated by this one receiver. 
Then, the updated counter for the integrity-protecting tag is broadcasted by a CAIBA-authenticated CAN frame to all receivers.
This procedure guarantees that no nonce is reused (for authentication by the sender), while even for counters that drifted apart significantly, they are reset to the same value.

\subsection{Reliability with Multiple Authenticators}
\label{sec:integration:multiple}

A single authenticator is a potential weakness of \name's design.
Such a single point of failure can interfere with high-reliability demands of applications relying on CAN.
While an inactive or misbehaving authenticator is quickly identified by the sending controllers~(\cf~\autoref{sec:integration:transmitter}), downgrading to unprotected communication should always be avoided.

Therefore, we propose to operate \name with multiple authenticators.
However, if many authenticators overwrite the same bit simultaneously, the physical signal is soon disturbed and reliable sensing is unlikely, especially due to different propagation delays.
Instead, authenticators should be distributed along the bus and only the closest authenticator takes care of overwriting a signal. 
Which authenticator is responsible for which \ac{ECU} can be preconfigured, as \acp{ECU} are usually stationary.
Such a multi-authenticator deployment has the advantage that the sender is always close to its dedicated authenticator, such that propagation delays are reduced.
If an authenticator is malfunctioning, it or a noticing \ac{ECU}, can inform the next authenticator in line to take over.
Thus, multiple authenticators increase reliability by (1) reducing the physical distance between senders and authenticators, and (2) offering fallback authenticators in case of malfunctions. 

\section{Fast MAC Scheme}
\label{sec:mac}

%In the following, we present the MAC scheme used for source-authenticating tags in \name.
For \name, we adapt the BP-MAC~\cite{2022_wagner_bpmac} scheme to allow the online computation of the final tag that can keep up with the speed requirements of the authenticator.
We first briefly recapitulate BP-MAC in \autoref{sec:mac:primer} (the original paper~\cite{2022_wagner_bpmac} provides more details), then show how we achieve online tag computability in \autoref{sec:mac:online} before discussing its integration into \name in \autoref{sec:mac:caiba}.

\subsection{A Primer on BP-MAC}
\label{sec:mac:primer}

\begin{figure}
    \centering
    \includegraphics[width=\columnwidth]{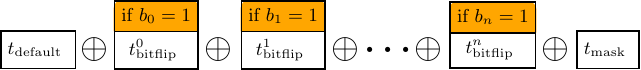}
    \vspace*{0.05cm}
    \caption{
    A BP-MAC tag is computed by XORing a default tag $t_{\text{default}}$, a masking tag $t_{\text{mask}}$, and a bitflip $t_{\text{bitflip}}$ for each bit~$b_i$ in a message that is set to 1. All of these individual tags can be computed ahead of time, only the XOR needs to be performed once the message is known.}
    \label{fig:bpmac}
\end{figure}

BP-MAC~\cite{2022_wagner_bpmac} is based on the Carter-Wegman MAC construction and optimized for short messages that are only a few bytes long.
An authentication tag is thus composed of a digest computed by a universal hash function over the message generated with an AES key $k_1$ and a masking tag in the form of a pseudo-random number generated with a distinct AES key $k_2$.
The masking tag is responsible for hiding the digest, as the MAC is only secure as long as no attacker learns any of these digests.
For BP-MAC's universal hash function, each bit is processed individually which allows the preprocessing of these results with linear space overhead.
Concretely, the tag computation in BP-MAC works as shown in \autoref{fig:bpmac}.
In advance, a default tag $t_\text{default}$ is computed by XORing the bit tags of a message only composed of zeros.
Here, bit tags are the AES-encrypted bit index and value, \ie the bit tag $t_i$ of the $i$-th bit $b_i$ is computed as $\text{AES}_{k}(i,b_i)$.
Then, depending on where bits in the actual authenticated message are one, the $t_\text{default}$ is XORed with bitflip tags, \ie $t^i_{\text{bitflip}}$ = $\text{AES}_{k_1}(i,0) \oplus \text{AES}_{k_1}(i,1)$.
Finally, the resulting tag is masked with a masking tag that is computed by AES-encrypting a counter that is incremented with each message, \ie $\text{AES}_{k_2}(\text{\emph{counter}})$.
The counter for the masking tags is known in advance and can thus be precomputed.
%Thus \jb{Zusammenhang mir unklar}\cut{, BP-MAC offers a quick tag computation for short messages and due to its Carter-Wegman MAC construction, it offers the same security guarantees as the underlying AES algorithm.}\jb{Wiederholung}

\begin{figure}
    \centering
    \includegraphics[width=0.92\columnwidth]{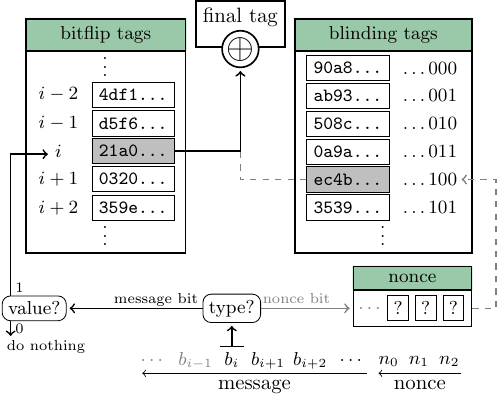}
    \caption{BP-MAC tags can be computed incrementally with each received bit, such that a single XOR operation suffices to compute the final tag once the final bit is read.
    }
    \label{fig:online-bpmac}
\end{figure}

\begin{figure*}
    \centering
    \begin{subfigure}[t]{.32\textwidth}
        \includegraphics[width=\textwidth]{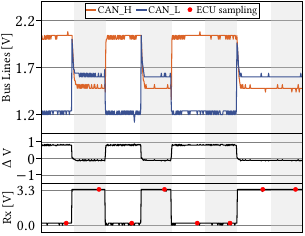}
        \caption{Unmodified transmission: \texttt{01010011}.}
        \label{fig:overwrite:normal}
    \end{subfigure}
    \hfill
    \begin{subfigure}[t]{.32\textwidth}
        \centering
        \includegraphics[width=\textwidth]{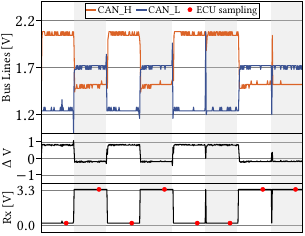}
        \caption{Original transmission \texttt{10101100} statically overwritten by \texttt{01010011}.}
        \label{fig:overwrite:direct}
    \end{subfigure}
    \hfill
    \begin{subfigure}[t]{.32\textwidth}
        \centering
        \includegraphics[width=\textwidth]{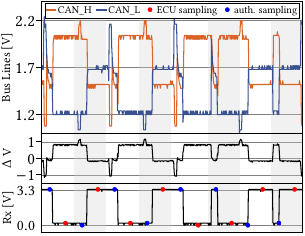}
        \caption{Reactively flipped original transmission of \texttt{10101100} to \texttt{01010011}.}
        \label{fig:overwrite:reactive}
    \end{subfigure}  

    \caption{Recording of the bus lines and transceiver's RX output during transmissions including sampling points.}
%    Showing CAN Low and CAN High for a the bitstream 01010011. In Figure~\ref{fig:overwrite:normal} the bitstream is directly written by the sender. In Figure~\ref{fig:overwrite:direct} and Figure~\ref{fig:overwrite:reactive}, the sender writes the bitstream 10101100 that is overwritten by the authenticator with prior knowledge of the sent bits or reactively, respectively.}
%        \label{fig:overwrite:reactive}
%    \end{subfigure}  
%
%    \caption{Showing CAN high and low for a the bitstream 0101\,\textbf{0011}. In Figure~\ref{fig:overwrite:normal}, the bitstream is directly written by the sender, whereas in Figure~\ref{fig:overwrite:direct} and Figure~\ref{fig:overwrite:reactive}, the bitstream 1010\,\textbf{1100} written by the sender that is overwritten by the authenticator with prior knowledge of the sent bits or reactively, respectively.}
    \label{fig:overwrite}
\end{figure*}

\subsection{Online BP-MAC Computations}
\label{sec:mac:online}

We specifically choose the BP-MAC scheme for the source-authenticating tag in \name 
since we can modify it for online tag computation.
Only with this online tag computation, the authenticator can operate fast enough to be ready to overwrite the next bit.
This adapted process to compute BP-MAC tags is illustrated in \autoref{fig:online-bpmac}.

The final tag variable is initialized with the default tag for a message composed only of zeros.
As we observe in the timeline at the bottom, data is then received bit by bit, and currently the bit $b_i$ of the message is processed.
Its value is checked and if the bit is set, the corresponding bitflip tag is XORed with the final tag.
These bitflip tags are precomputed and stored alongside the key for the corresponding transmission source.
Nonce bits are either explicitly transmitted or implicitly tracked by the sender and receiver.
If transmitted, the~(least-significant) nonce bits can be received and processed before or after the message bits.
In \autoref{fig:online-bpmac}, we show an example where the last three bits $n_0$, $n_1$, and $n_2$ are updated with the transmission, while the rest of the expected nonce is tracked implicitly.
This update process allows self-synchronization in case of a short burst of failed transmissions.
Once the complete nonce is known, the corresponding masking tag is selected and XORed with the final tag.
If the message and nonce bits are processed, the order of which actually does not matter, the final tag is known.
In either order, the process after the last bit is received is composed of a simple conditional check and at most one XOR operation.
Thus, the %\name 
authenticator can compute source-authenticating tags over the payload at line rate and be ready to potentially flip the first bit of the transmitted tag.

\subsection{Integration into the \name Authenticator}
\label{sec:mac:caiba}

In \name, we use \SI{24}{\bit} tags and \SI{4}{\bit} for nonce synchronization, which corresponds to \code{SecOC Profile 3 (JASPAR)} in the AUTOSAR standard~\cite{autosar}.
The blinding tag in BP-MAC is computed by AES encrypting a counter.
In \name, we take advantage of the fact that only each fifth blinding tag requires an invocation of the AES algorithm.
A \SI{16}{byte} encrypted AES block is divided into five blinding tags of \SI{3}{bytes}~(\SI{1}{byte} is discarded).
Thus, the counter is only encrypted if it is dividable by five, otherwise, the already computed blinding tags are used in order.
This encryption of an AES block happens whenever the message using the last blinding tag is transmitted.
If the message frequency is too high for the authenticator's processor, the AES algorithms can also be computed iteratively between every single message.

\section{\name's Overwriting Mechanism}
\label{sec:bitflip}

Altering CAN messages during their transmission is one key function of the authenticator in \name.
%The whole overwriting mechanisms are shown in \autoref{fig:overwrite:concept}, whereas the red part is the relevant part of the physical overwriting.
Technically, it requires a modification of the voltage levels on the bus lines and has to be timed precisely to not disturb ongoing transmissions and to ensure the desired message is received by other participants.
While the authenticator itself does not verify the transmitted tag $t$~($t = t^{i} \oplus t^{s}$), it calculates the source-authenticating tag $t^{s}$ based on the payload, the nonce, and the known key \ksource.
By then applying the XOR operation on $t$ and $t^{s}$~(recomputed based on the received data) again within the transmission, the resulting tag $t^{i}$ is read by the other devices of the bus.
Recently, the feasibility of real-time perfect bit modification attacks, where the original bus state is overwritten with a static value, has been demonstrated~\cite{2023_yousik_bit_modification}.
However, neither $t$ nor $t^i$ is known to the authenticator in advance, %before the message is received, 
such that the authenticator has to carry out these bit modification operation \textit{reactively} and in a bit-wise manner. 
Specifically, the authenticator must write the inverse bit signal of what was originally written to the bus whenever a bit in $t^i$ is set.
%, bit-per-bit as the bus lines take the according voltage levels.
%%The authenticator uses it to verify \jb{stimmt das? letztlich kann er ja ohne den group key nicht verifizieren, oder?} the source-authenticating tag $t^{s}$ by performing an XOR operation with the transmitted authentication tag $t$.
%%Due to the strict separation of knowledge about the source key \ksource and the group key \kgroup \jb{diese klare Aussage hat mir in Sec. 4 gefehlt.}, $t$ is not known to the authenticator. 
%Instead, the authenticator has to perform the signal modification \textit{reactive}, at which each bit signal of $t$ has to be read from the bus first, before the decision can be made to flip the bit.
In the following, we will describe the process of overwriting the authentication tag of a message in \autoref{sec:bitflip:physical}.
With \ac{RBF}, we then introduce a novel technique to reactively change the physical CAN signals on-the-fly in \autoref{sec:bitflip:rbf}.
Even though we use this technique exclusively for the \name authenticator, it is worth to emphasize that \ac{RBF} can be used more generally, for both defensive and offensive purposes.

\subsection{Physical Signal Modification}
\label{sec:bitflip:physical}

Flipping a single bit signal requires modifying voltage levels on the bus lines.
Overwriting a recessive bit to a dominant one is a core functionality of CAN to realize arbitration.
Hence, the authenticator can use the normal process to write a dominant bit and thus overwrite the recessive bit of another \ac{ECU}. %, resulting from the wired-AND circuit of the transceivers~(\cf~\autoref{sec:bg:dll}).
%Whenever a logical zero has to be written, the transmission of a dominant bit is enabled in a regular transceiver, resulting in a current flow trough its transistors and therefore the voltages on the bus lines reach their dominant level.
%is part of the arbitration and acknowledge method of CAN and is possible to any node due to the physical properties of the bit signals using a regular transceiver.
%We make use of it at the authenticator node to flip recessive bits to dominant during the transmission of the authentication tag $t$ by the sender.
%Overwriting dominant signals is not intended to be done in CAN and requires additional hardware.
%
%However, if the original transmitter is transmitting a dominant bit, the authenticator must actively drive the voltages on the bus lines to the recessive state to overwrite with a recessive bit.
%To achieve this overwriting, the current flowing through the original sender's transistors has to be sourced~(in the case of $CAN\_L$) or sunk~($CAN\_H$) by the overwriting transceiver.
%Additional components, which effectively erases the voltage applied by the original transceiver would typically need to have an output impedance that is significantly lower than the original transceivers' output impedance.
%However, as CAN is always operating based on the differential voltage, it is enough to keep the difference between the voltage levels of both lines below \SI{0.5}{\volt} to make sure it is read as a recessive state~\cite{iso11898-2}.
However, if the transmitter sends a dominant bit, the authenticator must actively drive the voltages on the bus lines to the recessive state to overwrite with a recessive bit.
To achieve this overwriting, the applied voltages of the transmitter have to be reverted by sourcing CAN\_L while sinking CAN\_H to ground.
This inverse connection, in relation to the transmitting transceiver, enables the current flow from the transceiver through the terminating resistors to be diverted off the bus, which reduces the measurable voltage drop of dominant signals.
However, as CAN always operates based on the differential voltage, it is enough to keep the voltage drop below \SI{0.5}{\volt} to make sure it is read as a recessive state~\cite{iso11898-2}.
Thus, additional components, which effectively erase the voltage difference would typically need to have an internal impedance that is significantly lower than the effective impedance of both terminating resistors.

\begin{figure}
    \centering
    \includegraphics[width=\columnwidth]{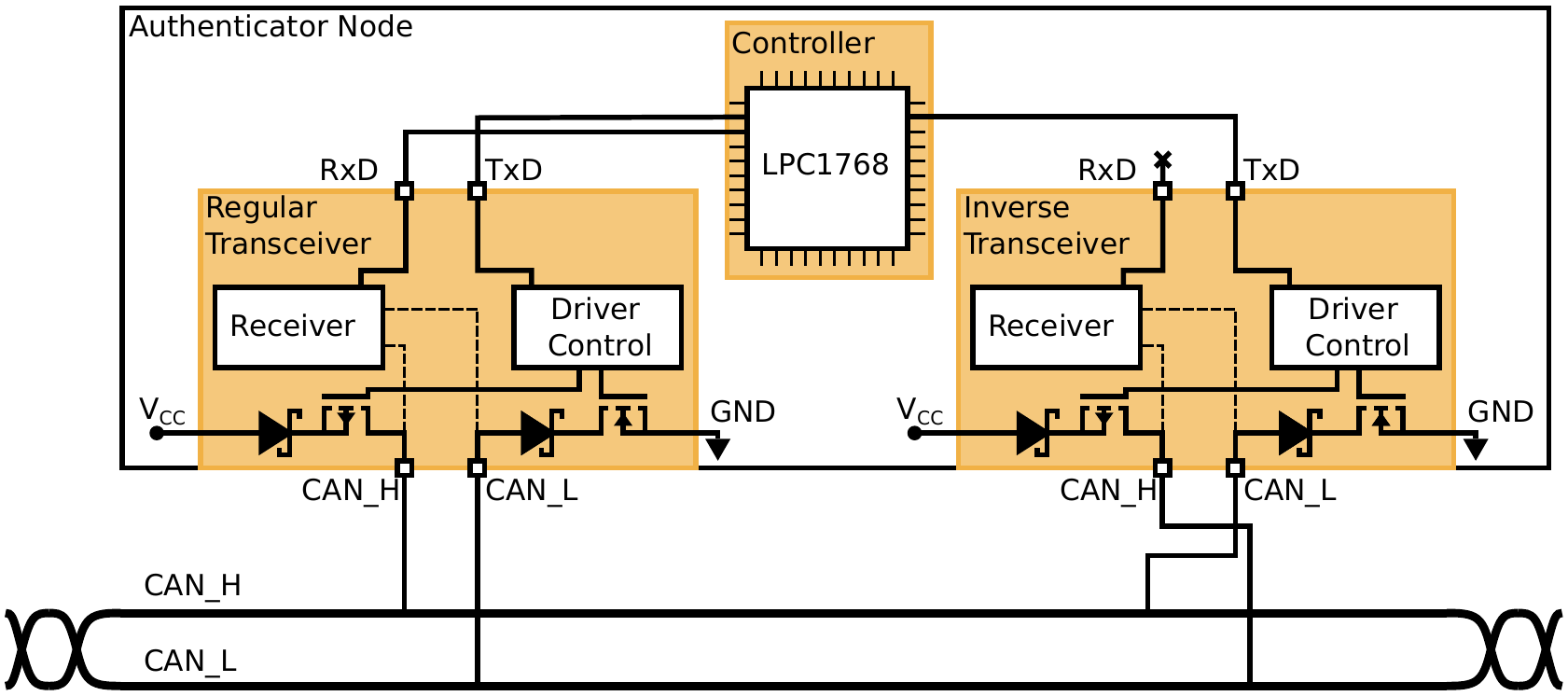}
    \caption{The authenticator node controls an additional CAN transceiver, that is inversely connected to the bus lines to \emph{erase} voltage differences between the bus lines.}
    \label{fig:overwriting}
\end{figure}

We could build such a device from discrete components.
However, ordinary CAN transceivers often already offer the inverted functionality.
A common internal structure of the transceiver contains two transistors to pull CAN\_H to the supply voltage and CAN\_L to ground~\cite{2016_griffith_transceiver}.
We can thus connect an additional CAN transceiver with inverted bus lines to the authenticator, as shown in \autoref{fig:overwriting}.
Then, we use the transistors to reduce the voltage difference when transmitting an apparent zero bit at this transceiver.
As this would typically result in a dominant transmission, it enables a current flow through the internal transistors, which sinks the current from CAN\_H to ground and sources CAN\_L from the supply voltage. 
This changes the bus voltage to the original, recessive state as can be seen in \autoref{fig:overwrite:normal} and~\subref{fig:overwrite:direct}.

\subsection{Reactive Bit Flipping}
\label{sec:bitflip:rbf}
The process to flip a bit depends on the current bit of the source-authenticating tag $t^{s}$ and the transmitted authentication tag $t$.
While $t^{s}$ is known, the authenticator, since not in possession of \kgroup, cannot compute a single bit of $t^{i}$ before the according bit of $t$ has been transmitted by the original transmitter.
%However, calculating this bit after receipt by the receiver which expects the tag to be $t^{i}$ to be correctly authenticated would result in a failing verification.
However, calculating this bit after sampling the bus would result in a failing verification, since the receiver would already expect the authenticated corresponding bit of $t^{i}$.
%Resulting from these requirements
Hence, signal modifications, as described in the previous  \autoref{sec:bitflip:physical}, requires a priori knowledge about the current bit signal. %, in order to be able to reactively overwrite bits.

Our proposed \ac{RBF} technique makes the overwriting mechanism of the authenticator \textit{reactive}, \ie the authenticator simultaneously reads and reacts to the current state of the bus within the transmission of the same bit. %, as shown in \autoref{fig:overwrite:concept}.
To be able to read the original signal, we modify the transmission only between the third and the last time quantum of each bit signal.
The first two time quanta are used to let the bus enter the state of the originally transmitted bit.
At the end of the second time quantum, the authenticator samples the bus and reads the latest transmitted bit of $t$~(\cf~\autoref{fig:overwrite:reactive}).
%This can be seen in \autoref{fig:overwrite:reactive} as the authenticator's sample point.
% We can expect that at this time, the bus is collision-free because it was reserved during the arbitration phase of the frame~(\cf~\autoref{sec:bg:dll}).
We can expect the bus to be collision free at this time, because it was reserved to the sender during the arbitration phase of the frame.
Furthermore, we can assume that the signal has already propagated via the bus due to an approximately constant propagation delay and shift of the bit timings between \acp{ECU}.
% Thus, the signal at the second time quantum is used for the bit flipping decision and to select the suitable CAN transceiver~(regular or inverse, \cf~\autoref{fig:overwriting}) for a possible bit alternation.
The sampled signal is then used for calculating the corresponding bit of $t^{i}$ and to select the suitable CAN transceiver~(\ie regular or inverse~(\cf~\autoref{fig:overwriting}) for a possible bit alternation.

In the remaining time quanta, the bus is forced into the desired state by the authenticator.
After \SI{75}{\%} of the nominal bit time has passed, \acp{ECU} sample the bus~\cite{canopen} and will read the authenticated bit of $t^{i}$.
% After the last time quantum of the received bit signal, this transceiver is reset to a recessive state again.
% The receiving \acp{ECU} only sample the bus after \SI{75}{\%} of the nominal bit time have passed~\cite{canopen}, such that the overwritten signal has time to propagate to each node.

\subsubsection{Bit Synchronization Conflict}

Changing the voltage levels on the bus within a bit time can conflict with CAN's edge-oriented synchronization.
Due to \ac{RBF}, bits are overwritten in the third time quantum which introduces additional edges.
%We have found one situation in which altering the bus state causes an additional edge that is used for resynchronization. %, \chk{which can be seen after the x-th bit in Figure~\ref{fig:overwrite:dominant}}. \jb{Gäbe es dazu einen Plot?}
Concretely, a \ac{CAN} controller synchronizes based on the edge when changing from a recessive state to a dominant state by prolonging the expected duration of the bit by the preconfigured Synchronization Jump Width~(SJW)~\cite{iso11898}.
%The SJW is 1 to 4 time quanta long and defines the maximum time a controller extends or shortens a bit by, where a larger number is generally chosen to improve robustness.
The SJW is 1 to 4 time quanta long and defines the maximum time by which a controller extends/shortens a bit, with a larger number generally chosen to improve robustness. 
However, altering the transmitted recessive bit to dominant will cause an edge in the third time quantum.
% Exactly if $t=1$ is followed by $t^i=0$ and $t^s=1$, the first detected edge occurs when the authenticator changes a signal from recessive to dominant, and all nodes resynchronize their bit time by this delayed edge. 
Exactly if the $j$-th bit is read recessive ($t^{i}_j=1$) and is followed by a dominant bit that is flipped ($t_{j+1}=1$ and $t^i_{j+1}=0$), the first detected edge occurs when the authenticator changes a signal from recessive to dominant, and all nodes resynchronize their bit time by this delayed edge. 
We compensate for the time shift by increasing the current bit time at the authenticator by two time quanta.
The increased time the authenticator overwrites the transmission ensures that the bit value remains constant until all nodes have sampled the bus.
Furthermore, we avoid additional disturbances to the bus, caused by multiple signal changes in a short period.
Thus, \name's authenticator compensates for the synchronization procedure embedded into all \ac{CAN} ECUs.

\subsubsection{Bit Stuffing Conflict}

Most parts of a CAN frame are affected by bit stuffing, including the authentication tag $t$ within the payload. 
Changing single bits in $t$ can require additional stuff bits or make existing ones obsolete.
Both cases could lead to an incorrect authentication tag received by other nodes or an incorrect length of the data frame and would interrupt the transmission.
As stated in \autoref{sec:integration:transmitter}, we expect the sender to place stuff bits according to the modified bit stream that is received.
%A possible different implementation would be to precompute the places of stuff bits using the group key \kgroup, which would however require a change to the transmitter's MAC.
To prevent overwriting stuff bits in $t$, the authenticator pauses the bit modification for one bit time whenever it expects a stuff bit from the sender.
The position of a stuff bit is determined based on the last five regular sampling points. %\cut{ and the assumption that the sender uses a correct placement of stuff bits}.
Although bit modification is paused, the authenticator samples the bus during the expected stuff bit to detect stuff errors or error frames.

\section{Evaluation}
\label{sec:eval}

We presented \name as an innovative source authentication scheme to protect CAN without requiring significant changes to existing systems.
In the following, we present a proof-of-concept implementation and show the general applicability of \name.
We start with the introduction of our evaluation setup and limitations in \autoref{sec:eval:setup}.
In \autoref{sec:eval:reliability}, we compare the reliability of our scheme regarding \name-protected traffic.
\name's compatibility with legacy CAN devices is demonstrated in \autoref{sec:eval:compatability}, its processing overhead investigated in \autoref{sec:eval:processing}, and potential bus length restrictions discussed in \autoref{sec:eval:length}.
Finally, we take a look at potential adverse long-term effects on the employed hardware in \autoref{sec:eval:longterm}.

\subsection{Evaluation Setup and Limitations}
\label{sec:eval:setup}

\begin{figure}[t]
    \centering
    {%
    \setlength{\fboxsep}{-.5pt}%
    \setlength{\fboxrule}{.5pt}%
    \fbox{\includegraphics[width=\columnwidth]{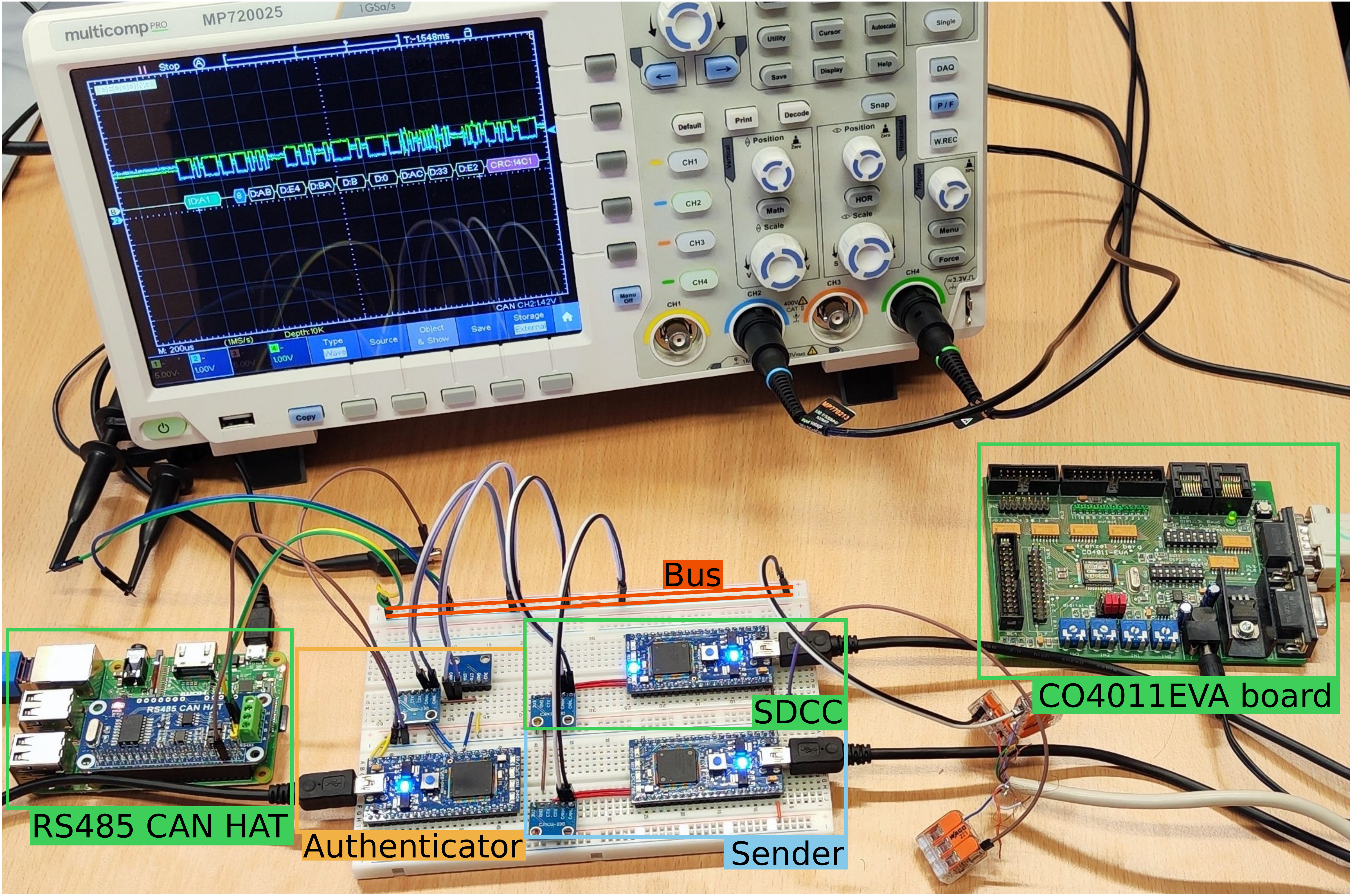}}%
    }%
    \caption{Evaluation setup of \name consisting of a \name-capable transmitter \textcolor{sender}{\textbullet\,}, three unmodified receivers \textcolor{receiver}{\textbullet}, and a \name authenticator \textcolor{authenticator}{\textbullet}.}
    \label{fig:setup}
\end{figure}

For our proof-of-concept implementation, we set up a testbed  consisting of different \acp{ECU} as shown in \autoref{fig:setup}.
To rapidly prototype the necessary modifications within the CAN controller, we used the software-defined CAN controller~(SDCC)~\cite{2019_cena_sdcc} with the recommended NXP~LPC1768 microcontroller platform and TI~SN65HVD230 CAN transceivers.
Additionally, three unmodified off-the-shelf CAN controllers were used to evaluate the backward compatibility of our solution.
%\jb{Controller besser hier vorstellen als in Sec. 8.3?} Nein zuviel Detail für die Übersicht
As the SDCC is a pure software implementation on a regular microcontroller without any hardware acceleration, the maximum bitrate is limited to \SI{40}{kbit/s}~\cite{2019_cena_sdcc}.
To show the general feasibility of \name, this is sufficient. 

However, to evaluate \name with higher bitrates, a hardware- or FPGA-based implementation would be necessary.
This is an effort we consider disproportionate %for showing the general feasibility of \name.
to demonstrate \name's general feasibility.
Instead, our evaluation focuses on the practical feasibility %of \name 
at low bitrates, while we theoretically analyze the effects of higher bitrates \eg \wrt to the bus length.
However, as commercial adaptions of \name are expected to be implemented in hardware, \eg as SIP cores~(\cf~\autoref{sec:integration:deployment}), we do not anticipate any processing %or timing 
limitations even at higher speeds.
For example, the most time-critical aspect is the final computation of $t^s$ by the authenticator after the last payload bit has been received, which requires a single XOR operation of three bytes and can be computed within a single clock cycle by a hardware-based authenticator. 

With a faster hardware prototype, we could validate that \name and legacy CAN ECUs can coexist on the same bus in an actual car.
However, even then, we could not easily investigate if receiving ECUs implementing the AUTOSAR SecOC standard could be retrofitted with \name's source authentication because we have no access to the secret group key used by the car's ECUs.
Consequently, we have to fall back to a physical testbed for our proof of concept evaluation of \name.

%\subsection{Why Not Validate \name in a Car?}

%One could ask if it would not be worth it to design a faster \name prototyp simply to evaluate \name in an actual car, usually operated at CAN speeds of \SI{500}{kbit/s}.
%However, this would not be possible.
%If the bus would not support AUTOSAR SecOC, \name ECUs would simply fallback to standard CAN.
%On the other hand, if ECUs would support the AUTOSAR SecOC standard, \name frames would still be rejected for the simple fact that we do not have access to the secret group key.
%Hence, for now, we have to fallback to a physical testbed to evaluate \name.

\subsection{Reliability}
\label{sec:eval:reliability}

\begin{figure}[t]
    \centering
    \includegraphics[width=\columnwidth]{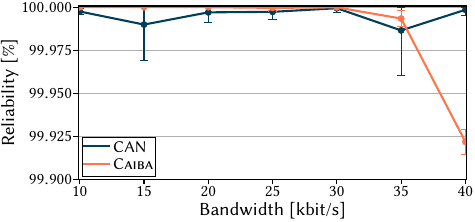}
    \caption{ \name achieves similar reliability to CAN for low data rates but it reaches the timing limitations of the SDCC slightly earlier when increasing the rate. }
    \label{fig:reliability}
\end{figure}

Ideally, \name can enable source authentication to CAN without impacting its reliability.
Therefore, we compare the reliability of \name to a regular CAN deployment.
For data rates varying between 10 and 40\,kbit/s, we sent 100,000~frames with 4 to 8 bytes of payload~(incl. authentication tag) and analyze the ratio of correctly received frames at the receiver, including a valid integrity-protecting tag.
To have an equal number of \acp{ECU} connected to the bus, and thus have comparable desynchronization potential, we connect an additional listening \ac{ECU} for the CAN measurements that does not need the authenticator.
We repeat each measurement ten times and show our results~(incl. $95\,\%$ confidence intervals) in~\autoref{fig:reliability}.

For low data rates, we surprisingly observe that \name achieves slightly higher reliability than CAN with no frame loss observed for data rates below 35\,kbit/s.
The most likely explanation for this effect is that the authenticator \ac{ECU}, which acts as an additional CAN receiver for these measurements, leads to rare misinterpretations of the ACK bit.
As we reach the maximum data rates supported reliably by the SDCC~(\ie 40\,kbit/s), \name's reliability starts to decrease.
Here, we reach the limitations of the timing accuracy achievable in a pure software-defined controller slightly earlier than for CAN due to the delicate bit synchronization necessary during \ac{RBF}.
A similar drop in reliability can also be observed in CAN when increasing the data rates further. %\jb{Warum zeigen wir das dann nicht?} 

Overall, \name can operate with similar reliability as CAN.
This high reliability can however only be achieved with an authenticator providing precise timing relative to the bus speed.
For higher data rates, this is only achievable through hardware-implementations of \name controllers.

\subsection{Compatibility with Legacy CAN Devices}
\label{sec:eval:compatability}

%With regard to incremental retrofitting, 
We cannot expect that each device on a bus is aware of \name.
Otherwise, car manufacturers would have to convince each ECU supplier to adopt \name, before the first car can employ multicast source authentication.
Thus, \name is specifically designed to not interfere with legacy communication.
Also, only transmitters have to be altered to support \name, while AUTOSAR SecOC-implementing receivers can still benefit from the provided security without any modifications~(\cf~\autoref{sec:integration}).

To validate the compatibility with legacy devices, we tested \name in combination with three additional off-the-shelf receivers. % beyond the SDCC running on the LPC1768 boards.
In particular, we used \name in combination with an MCP2515 CAN controller connected to a Raspberry\,Pi\,3 trough SPI in the form of the \textit{Waveshare\,RS485 CAN HAT}~\cite{canhat}, the integrated CAN interpreter of a \textit{MULTICOMP\,PRO MP720025\,EU-UK}~\cite{multicomp} oscilloscope, and a CANOpen Evaluation Board using the \textit{Frenzel+Berg CO4011A %CanOpen 
Controller}~\cite{co4011}.
All three devices served as regular receivers for authenticated CAN messages that were modified by the \name authenticator.
As a result, no anomalies have been observed when reading CAN messages with any device.
Thus, we conclude that \name is indeed compatible with unmodified receivers and legacy CAN communication. 

\subsection{Upper Bound on Processing Overhead}
\label{sec:eval:processing}

In the literature, cryptographic approaches to protect the CAN bus are often criticized for excessive processing overhead.
Indeed, the computation of a single HMAC-SHA256 tag on an ARM\,Cortex\,M3~(32\,MHz) takes $1641.4 \pm 0.1 \mu s$.
The ARM\,Cortex\,M3 is a typical prototyping processor with similar processing power to commercial \acp{ECU}, which are however built to operate reliably even in harsh environments.
The computation of one tag thus takes over 10 times longer than the transmission of a single CAN frame at \SI{1}{Mbit/s}~(\SI{128}{\mu s}).
In other words, a corresponding \ac{ECU} could only verify or authenticate one tenth of all CAN frames transmitted on a fully utilized \SI{1}{Mbit/s} bus.
To assess the overhead of \name on the \acp{ECU}, we measured the duration to compute a tag~(consisting of $t^s$ and $t^i$) with a length of \SI{24}{bit} for payload lengths varying between 1 and 5 bytes.

\begin{figure}[t]
    \centering
    \includegraphics[width=\columnwidth]{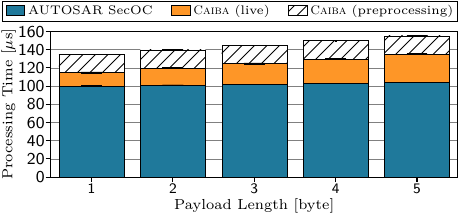}
    \caption{Processing overhead of \name's tag computation on a representative ARM Cortex M3 chip is significantly lower than a single frame transmission~(\SI{128}{\mu s}) even if CAN is operated at \SI{1}{Mbit/s}.}
    \label{fig:latency}
\end{figure}

For the integrity-protecting tag $t^i$, we compute a CMAC tag as recommended in AUTOSAR SecOC. 
On the other hand, the source-authenticating tag $t^s$ is computed based on BP-MAC for fast online computation by the authenticator.
As our results in \autoref{fig:latency} demonstrate, the overhead of the additional tag computation is at most \SI{51}{\%} in total.
Here, a fixed overhead of \SI{19.9}{\mu s} is caused by the preprocessed computation of blinding tags.
The actual additional delay during tag computation only amounts to between \SI{14.6}{\mu s} and \SI{31.3}{\mu s}, depending on the payload length.
Moreover, we observe that if AUTOSAR SecOC were to adopt BP-MAC for its tag computation, it could reduce its processing time by \SI{69.0}{\mu s}, \ie \name would then be faster than AUTOSAR SecOC with its recommended MAC scheme of today.

As we expect that commercially deployed \name controllers are implemented in hardware, the overhead of BP-MAC is suspected to further reduce significantly~\cite{2022_wagner_bpmac}.
Overall, we found that cryptographic processing is not a significant drawback for \acp{ECU}, even for cryptographic processing in software, due to the selection of a fast, yet secure, MAC scheme.

\subsection{No Limitations to the Bus Length}
\label{sec:eval:length}

As the authenticator's overwritten signal must be reliably sampled by the receiving \acp{ECU}, we now look at potential bus length restrictions.
In the worst case, the originally transmitted signal and the overwritten signal are delayed by twice the distance between the sender and the authenticator.
This maximal offset occurs if the receiver is placed near the sender on one extremity of the bus with a maximum distance to the centrally placed authenticator.
Hence, one might expect a reduction in the maximum CAN cable length by the employment of \name.
However, adding up the delays for signal propagation, overwriting, and synchronization even for the maximal bus lengths, \name's delays are still tolerable according to the acceptable intervals for all bit rates of % supported by 
CAN~(assuming a single centrally placed authenticator).

Looking at the example for \SI{1}{Mbit/s}, CAN supports a maximum cable length of \SI{25}{m} and a receiver sampling point \SI{750}{ns} after the start of a bit.
Adding the propagation delay of traveling halfway and back~(\SI{125}{ns}), three quanta idle time for synchronization and back-propagation~(\SI{375}{ns}), and the authenticator's transceiver delay~(\SI{210}{ns}) results in a worst-case delay of \SI{690}{ns}\footnote{Baseline numbers stem from the CANopen standard~\cite{canopen}.}.
Hence, even with a maximum CAN cable length and worst-case sender and receiver placement, the overwritten signal is still stable at the receiver before it samples the bus \SI{750}{ns} after the start of the signal.
Using \name thus does not restrict maximum cable length, as other aspects of CAN are more restrictive, \eg arbitration and ACK bits that must function over the entire length of the bus. 
Meanwhile, the centrally placed authenticator is closer to the sender and thus operates with lower propagation delay.

To practically validate that \name can operate on longer buses, we placed a \SI{50}{\meter} twisted pair cable between the authenticator on the one end of the cable, and the sending as well as the receiving \ac{ECU} on the other end of the cable.
This corresponds to the worst-case scenario for a \SI{100}{\meter} long bus.
We transmit 10,000 CAN frames at \SI{40}{kbit/s} and repeat this measurement ten times.
Note that at these speeds \ac{CAN} could operate on a bus 10 times as long, but only reliably with the use of optocouplers~\cite{canopen}. 
We achieve a reliability of $99.95 \pm 0.03$\,\%.
While this reliability is even slightly higher than for the short bus ($99.92 \pm 0.01$\,\%), potentially due to better stabilizing of the signal during propagation, they lie within the margin of error of each other.
Overall, we can thus conclude that \name can operate reliably even on longer buses and does, in theory, not restrict the maximum bus length at all.

\subsection{Long-Term Impact of Overwriting Bits}
\label{sec:eval:longterm}

Finally, we study potential long-term adverse effects of \name{} on the hardware.
When the authenticator overwrites a dominant bit, the inversely connected transceiver sinks the current from the bus line, resulting in an actively driven recessive state. 
While this can be seen as a short circuit on the bus (from the view of the original transceiver) with a dampened current,
 all transceivers must be short-circuit proof for fault tolerance~\cite{iso11898-3}.

Additionally, we measure the current that flows through the affected transceivers during the overwriting of dominant bits.
We observe a maximum current of \SI{28}{mA}, which is within specifications for common CAN transceivers such as the Philips TCA1050~(\SI{100}{mA})~\cite{tja1050} or the TI SN65HVD25x~(\SI{160}{mA})~\cite{sn65}. 
Most importantly, these currents only affect \name transceivers. 
Legacy \acp{ECU} connected to the same bus are not affected as they only listen when dominant bits are overwritten.
Consequently, there exists no risk for long-term effects of operating \name in \ac{CAN}.

\section{Limitations and Future Challenges}

We present \name as a novel solution to the multicast source authentication problem and thus protect CAN.
In this context, we provide a proof-of-concept implementation and show its compatibility and reliance in a range of evaluations.
In the following, we identify current limitations and potential future improvements for \name.

\textbf{Broader Applicability of \name.}
\name promises multicast source authentication without verification delay or excessive bandwidth requirements.
Indeed, we demonstrate \name's seamless integration into the CAN bus protocol.
However, it remains to be investigated how widely applicable \name is beyond CAN.
Other automotive buses, such as LIN~\cite{lin} and FlexRay~\cite{flexray}, show great potential but also pose some unique challenges.
The potential of \name in these and other networks, such as star topologies with the central switch taking on the role of the authenticator, is yet unclear.\vspace{1mm}

\textbf{Hardware-based \name \acp{ECU}.}
We prototypically show the applicability of \name on a software-defined CAN controller and its interoperability with off-the-shelf CAN controllers.
However, inherent performance limitations and timing inaccuracies of software-defined controllers limit the bus speeds.
A hardware-based \name controller is thus needed for the deployment of \name with typical  bus speeds in cars.
\vspace{1mm}

\textbf{\name alongside Intrusion Detection.}
At the beginning of this paper, we argued that we should not rely solely on \ac{IDS} because of their imperfect detection, false alarms, and potential evasions.
Nonetheless, \acp{IDS} do not become superfluous because of \name.
In contrast, the operation of \name produces characteristic behavior that can even facilitate monitoring by an adapted intrusion detection mechanism to detect any anomalous behavior early.
Such potential symbiosis should be further investigated.\vspace{1mm}

\textbf{Hardening the Authenticator Module.}
Throughout this paper, we assume a trusted authenticator that is hard to compromise even with physical access to the CAN bus.
This assumption is in line with other proposals, \eg IDSs for in-vehicular communication.
Still, the concrete steps to maximize temper-resilience beyond offering no external connectivity must be worked out.
\vspace{1mm}

\section{Ethics Considerations} 

By proposing a preventive measure to secure communication on the CAN bus, our research might not raise obvious major ethical concerns.
Still, during the conceptualization and execution of our research we carefully followed established best-practices and guidelines to identify and address any existing ethics-related concerns~\cite{menlo_report,kohno2023ethical}.

Our survey and especially \hyperref[app:cryptanalysis]{Appendix~\ref{app:cryptanalysis}} summarize the weaknesses of existing CAN security mechanisms proposed in research.
The information summarized there could potentially be exploited by malicious actors.
However, our analysis is based on information that is mostly already publicly available and, with the exception of AUTOSAR SecOC~\cite{autosar}, we are not aware that any of these proposed schemes are used in commercial products.
For the AUTOSAR SecOC specification, the vulnerability to masquerading attacks is a known and accepted risk~\cite{autosar}.
Thus, we carefully ensured to not give malicious actors any advantage to compromise any CAN-controlled system through the information we provide in this paper.
In contrast, we raise awareness to weaknesses in previously proposed schemes, thus potentially preventing the deployment of vulnerable security schemes which could offer a false sense of security.

Concerning publishing the idea of \name, we may give malicious actors early access to understand the mechanism behind a security scheme that may be later deployed in cars or military vehicles.
Thus, malicious actors have more time to identify weaknesses and prepare for practically exploiting them.  
However, exposing security solutions to public scrutiny is important, such that the research community can jointly identify weaknesses before they can be exploited to derive overall more secure solutions.

\section{Conclusion}

The increasing connectivity of modern vehicles, coupled with the vulnerability of the \ac{CAN} bus, has raised significant concerns regarding the safety and security of automotive systems.
Compromised \acp{ECU} of, \eg infotainment systems, can easily masquerade as other entities in the network and take over critical control of a vehicle.
Current security mechanism mostly rely on intrusion detection or group key-based authentication tags, with both approaches not protecting sufficiently against masquerading attacks.
In contrast, we propose a novel multicast source authentication scheme (\name) for buses that, unlike prior approaches, does not require longer tags or delayed verification.
\name relies on an authenticator that reactively overwrites bits of the authentication tag included in each message.
The keys to generate the original tags are only known by the genuine source, while all receivers can verify the altered tags.
We show the applicability of \name in CAN, providing source authentication for \acp{ECU} implementing the AUTOSAR SecOC specification.
\name is incrementally deployable and interoperable with legacy CAN traffic, while achieving high reliability with minimal processing overhead.